\documentclass{jfm}

\usepackage{natbib}
\usepackage{bm}
\usepackage{amsmath}
\usepackage{graphicx,color}
\usepackage{subcaption}
\usepackage{tikz}
\usepackage{xcolor}
\usepackage{upmath}
\usetikzlibrary{arrows,calc}
\usepackage{multirow}
\usepackage{soul}
\usepackage{epstopdf}
\ifCUPmtlplainloaded \else
  \checkfont{eurm10}
  \iffontfound
    \IfFileExists{upmath.sty}
      {\typeout{^^JFound AMS Euler Roman fonts on the system,
                   using the 'upmath' package.^^J}%
       \usepackage{upmath}}
      {\typeout{^^JFound AMS Euler Roman fonts on the system, but you
                   dont seem to have the}%
       \typeout{'upmath' package installed. JFM.cls can take advantage
                 of these fonts,^^Jif you use 'upmath' package.^^J}%
      }
  \else
  \fi
\fi

\ifCUPmtlplainloaded \else
  \checkfont{msam10}
  \iffontfound
    \IfFileExists{amssymb.sty}
      {\typeout{^^JFound AMS Symbol fonts on the system, using the
                'amssymb' package.^^J}%
       \usepackage{amssymb}%
       \let\le=\leqslant  \let\leq=\leqslant
       \let\ge=\geqslant  \let\geq=\geqslant
      }{}
  \fi
\fi

\ifCUPmtlplainloaded \else
  \IfFileExists{amsbsy.sty}
    {\typeout{^^JFound the 'amsbsy' package on the system, using it.^^J}%
     \usepackage{amsbsy}}
    {\providecommand\boldsymbol[1]{\mbox{\boldmath $##1$}}}
\fi

 \newcommand\im{\ensuremath{\mathrm{i}}} %imaginary unit
\newcommand{\Vhat}[1]{{\widehat {\bm #1}}}

\providecommand\bcdot{\boldsymbol{\cdot}}

\newcommand\Real{\mbox{Re}} % cf plain TeX's \Re and Reynolds number
\newcommand\Imag{\mbox{Im}} % cf plain TeX's \Im
\newcommand\Pran{\mbox{\textit{Pr}}} % Prandtl number, cf TeX's \Pr product
\newsavebox{\astrutbox}
\sbox{\astrutbox}{\rule[-5pt]{0pt}{20pt}}

\providecommand\bcdot{\boldsymbol{\cdot}}

\newcommand{\bee}{\begin{equation}}
\newcommand{\ee}{\end{equation}}
\newcommand{\beal}{\begin{align}}
\newcommand{\eeal}{\end{align}}
\newcommand{\bea}{\begin{eqnarray}}
\newcommand{\eea}{\end{eqnarray}}

\providecommand\bcdot{\boldsymbol{\cdot}}

\definecolor{uc}{rgb}{0.679,0.2695,0.2539}
\definecolor{cr}{rgb}{0.3008,0.7188,0.7305}
\definecolor{so}{rgb}{0.5547,0.5391,0.7305}
\definecolor{bo}{rgb}{0.5781,0.3047,0.6953}
\definecolor{tr}{rgb}{0.2617,0.2617,0.8633}
\definecolor{lb}{rgb}{0.9961,0.6016,0.6172}
\definecolor{tb}{rgb}{0.7031,0.6211,0.5664}
\definecolor{sdp}{rgb}{0.5547,0.3789,0.3477}
\definecolor{lr}{rgb}{0,0.8164,0}
\definecolor{bi}{rgb}{0.03125,0.3477,0}
\title[Spatio-temporal Patterns in Inclined Layer Convection]{Spatio-temporal Patterns in Inclined Layer Convection}

\author[P. Subramanian \textit{et al.}]%
{Priya Subramanian$^{1,2}$%
  \thanks{Email address for correspondence: P.Subramanian@leeds.ac.uk}, \ns
Oliver Brausch$^3$, \ns Karen E. Daniels$^4$, \ns Eberhard Bodenschatz$^1$, \ns 
Tobias M. Schneider$^{1,5}$ \& Werner Pesch$^3$\thanks{Email address for
correspondence: werner.pesch@uni-bayreuth.de}}
\affiliation{$^1$Max-Planck Institute for Dynamics \& Self-Organization, G\"ottingen 37077, Germany\\
$^2$ School of Mathematics, University of Leeds, Leeds LS29JT, UK\\
$^3$ Universit\"at Bayreuth, Theoretische Physik I, Bayreuth 95447, Germany\\
$^4$Department of Physics, North Carolina State University, NC 27695, USA\\
$^5$Emergent Complexity in Physical Systems Laboratory (ECPS), {\'E}cole Polytechnique F{\'e}d{\'e}rale de Lausanne, CH-1015, Switzerland\\}
\date{\today; revised ?; accepted ?. - To be entered by editorial office}
\begin{document}

\maketitle

\begin{abstract}
This paper reports on a theoretical analysis of the rich variety of
spatio-temporal patterns observed recently in inclined layer convection at
medium Prandtl number when varying the inclination angle $\gamma$ and the Rayleigh
number $R$. The present numerical investigation of the inclined layer convection system is based on the standard Oberbeck-Boussinesq equations. The patterns are shown to originate from a complicated competition of 
buoyancy-driven and shear-flow driven pattern forming mechanisms. The former are expressed as 
\rm{longitudinal} convection rolls with their axes oriented parallel to the incline, the latter as perpendicular \rm{transverse} rolls. Along with conventional methods to study roll patterns and their 
stability, we employ direct numerical simulations in large
spatial domains, comparable with the experimental ones. As a result, we determine the phase diagram of the characteristic  complex 3D convection patterns
above onset of convection in the $\gamma-R$ plane, and find that it compares very well with the experiments. In particular we demonstrate that interactions of specific
Fourier modes, characterized by a resonant interaction of their wavevectors in
the layer plane, are key to understanding the pattern morphologies.
\end{abstract}

\section{Introduction}\label{sec:intro}
Pattern forming instabilities in macroscopic dissipative systems, driven out of
equilibrium by external stresses, are common in nature and have
been studied intensely over the last decades (see e.g. \cite{CroHo1993}).
Prominent examples   are  found in
fluid systems
(see  e.g. \cite{chandra1961}, \cite{swin85}) where the pattern formation is
driven either
thermally or by shear stresses. 
The general understanding of pattern forming systems has benefited
from numerous experimental and theoretical investigations of the
classical,
thermally driven
Rayleigh-B\'enard convection (RBC) in a layer of a simple fluid
heated from below \citep*{Busse1989,Bodenschatz2000,Lappa2009}. In the RBC system, the main control
parameter is the Rayleigh number, $R$, a dimensionless measure of the applied 
temperature gradient.
At a critical
Rayleigh number, $R_c$, the quiescent
heat-conducting basic state  develops into the well-known periodic arrays of
convection rolls characterized by a critical wavevector $\bm q_c$.
The stability of the rolls and their evolution towards characteristic  3D patterns via sequences
of bifurcations with increasing $R$ has been investigated by
Busse and
coworkers (see e.g. \cite{CroHo1993,Busball1996} and references therein). The present paper analyzes a variant of RBC, the inclined layer
convection (ILC) system, where  the 
fluid layer is inclined at  an angle $\gamma$ to the horizontal. Investigations
of this  system also have a long tradition
(see e.g. 
\cite{Vest1969,Gershuni1969,Hart1971,Bergholz1977,Ruth1980,Fujimura1992,Karen2000}).

%%%%%%% figure 1
\begin{figure}
\begin{center}
\includegraphics[width=10.5cm]{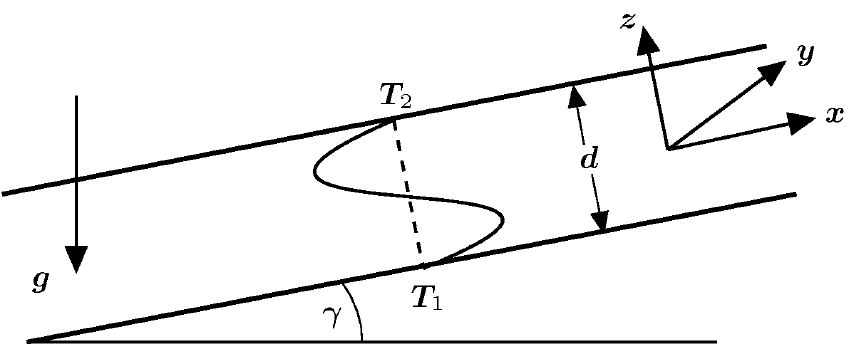}
\end{center}
\caption{Inclined convection cell of thickness $d$  
 which is heated from below and  cooled from
above with temperature difference $\Delta T \equiv T_{1} -
T_{2} >0$  for the inclination angle $0^{\circ} \le \gamma  \le 90^{\circ}$.
Driven by gravity $\bm g$ the
cold fluid flows downwards near the top plate and
the hot fluid flows upwards near the bottom plate in the form
of a cubic velocity 
profile  (\ref{eq:basest}). For the range $90^{\circ}
< \gamma < 180^{\circ}$, the fluid layer is inverted and is heated from above.} \label{fig:1}  
\end{figure}

In the ILC system, for $\gamma \ne 0^{\circ}$ gravity $\bm g$ has
components both  perpendicular and parallel to the fluid layer, which leads to
an important modification of the basic state compared to RBC.
The applied temperature gradient first produces stratified fluid layers
with continuously varying temperatures and densities.
In addition, the basic state already contains a flow field driven by the
in-plane component of $\bm g$: the heavier (colder)  fluid will flow 
down the incline and the lighter (warmer) fluid will flow upwards.
Since the resulting flow field creates a velocity gradient perpendicular
to the fluid layer, both buoyancy and
 shear stress driven instabilities of the basic state compete.  Their  relative
importance is governed by the Prandtl number $\Pran$, the ratio of the thermal
diffusivity, $\kappa$,  to the kinematic viscosity, $\nu$, of the fluid.
Furthermore, the strength of the shear stress can be continuously increased by increasing $\gamma$.
The orientation of the roll axes at onset of convection allows for directly discriminating between buoyancy and shear driving. The buoyancy driven rolls
are aligned parallel to the incline ({\rm {longitudinal rolls}}) while
the shear driven rolls are aligned perpendicular to the incline ({\rm{transverse
rolls}}). It should be noted that the latter also bifurcate, when the fluid
layer is  heated from
above and the thermal stress is therefore stabilizing.

Our goal is not  a representative  parameter study  of ILC with
respect to $R, \gamma, \Pran$, which would go beyond the scope of a single
paper. Our theoretical investigations have instead been motivated  by recent ILC
experiments in pressurized $\mathrm{CO_2}$ \citep{Karen2000,Bodenschatz2000}
with a fixed value $\Pran =1.07$  of the Prandtl number. In this work 
the $R-\gamma$ parameter space has been systematically explored 
and a variety of fascinating patterns have been described. As in all ILC studies mentioned above, our theoretical analysis is
based on the Oberbeck-Boussinesq equations (OBE). In contrast  to the
extensively studied  RBC, earlier results in the literature for the ILC system
are mostly limited to the linear regime and characterize the primary bifurcation
of the convection rolls from the basic state at $R = R_c$ (see e.g. \cite
{Vest1969,Gershuni1969,Hart1971,Ruth1980,Fujimura1992}). In the nonlinear regime ($R > R_c$) Busse and Clever \citep{Clever1977,Busse1992} investigated
secondary and tertiary instabilities of the convection rolls for some special
cases. In contrast, the present work is devoted to a comprehensive
theoretical analysis of the patterns in \citet{Karen2000} at  $\Pran =1.07$. 

In the present work we make use of the well-known arsenal of 
concepts to analyse pattern forming instabilities in fluid systems
(see e.g. \cite{CroHo1993}).  This approach deploys its full
power in large aspect ratio systems (lateral extension, $L$,  of the
fluid layer much larger than its thickness, $d$) which were first realized experimentally in 
\cite{Karen2000}. From a linear instability analysis of the basic state we determine the
critical values $R_c, \bm q_c$ at the onset of convection. The properties of the rolls in the weakly
nonlinear regime, $R \gtrsim R_c$,  are then analyzed  in the framework of amplitude
equations, which yield approximate roll solutions. Using these as starting solutions allows for the iterative determination of the roll solutions in the nonlinear regime, where $R >R_c$. In a further step, their stability is again tested  to identify the secondary instabilities of the roll pattern. 

We will demonstrate the agreement between experiments and theory with respect to the onset of convection in the $\gamma$-$R$-plane. For inclination angles $\gamma$ below a codimension $2$ angle $\gamma_{c2} \approx 78^{\circ}$ for $\Pran = 1.07$ the destabilisation of the basic state is driven by longitudinal rolls, while transverse rolls  bifurcate for $\gamma > \gamma_{c2}$. Both  bifurcations  are always stationary and continuous (supercritical). The subsequent secondary destabilisation of the  2D rolls  for $\gamma \ne 0^{\circ}$ at increasing $R$ is driven by {\it oblique} roll solutions, whose axes are not along the longitudinal or transverse roll directions. As a result, spatially periodic 3D patterns are often observed. These are characterized by the nonlinear interaction
of three roll modes with wave vectors ($\bm q_1,   \bm q_2\textrm{~and~}  \bm q_3$),  that
fulfil a wavevector resonance condition $\bm q_1 + \bm q_2 + \bm q_3 =0$.
 
As common in other large aspect ratio convection experiments,  one also finds imperfectly periodic, weakly turbulent patterns.  For instance, the 3D
motifs mentioned above appear locally superimposed on the original
the 2D roll pattern, where they  burst and vanish repeatedly in time
\citep{Karen2000,Karen2002,Karen2003}. To test whether such dynamic states are caused by  experimental imperfections (e.g. lateral boundaries, spatial variations of the cell thickness, inclination of the cell in two directions), we have performed the first comprehensive  numerical simulations of the OBE in ILC for large aspect ratio convection cells. While a conclusive assessment  of the underlying mechanism producing bursts remains elusive, the weakly turbulent dynamics of the pattern have been well reproduced in our simulations. Since the shear stresses play an important role  in our system, there might be an analogy to the instabilities of the laminar state in purely shear driven fluid systems like Couette or Poisseuille flow, which also often appear in the form of  localized events (for recent examples see e.g. \citep{Lemoult2014,Tuckerman2014}). We hope that investigations such as this one will reveal deeper commonalities between ILC and such shear driven patterns in the future.

The paper is structured as follows: a brief summary of the governing
OBE for the ILC system is given in \S\ref{sec:gde}. We then discuss the
onset of convection  in terms of $\bm q_c, R_c$ for $\Pran = 1.07$ and the resulting periodic
roll pattern in \S\ref{sec:rolls}. The results of the stability analysis
of the rolls   the nonlinear regime and the resulting phase diagram are
presented in \S\ref{sec:secstab}. In \S\ref{sec:comp} we show direct
simulations of the OBE for different $R$ and $\gamma$, which compare well with
the experiments. A short summary of this work  together with perspectives for future work can be found in \S\ref{sec:disc}. In three detailed appendices we present first the detailed OBE equations for the ILC system  and discuss  the numerical method to characterize the roll solutions above onset together with their secondary
instabilities (Appendix \ref{App:a}). Then we address  briefly our approach
to solve  the OBE in general    using direct numerical simulations (Appendix
\ref{App:B}). Finally we return to the linear stability analysis of the basic 
state to give additional informations regarding    the properties of $\bm q_c,
R_c$ (Appendix \ref{App:clin}).

\section{Oberbeck-Boussines equations for ILC}\label{sec:gde}
%%%%%%%%%%%%%%%%%%%%%%%%{\color{red} Equations and Oberbeck-Boussinesq

As shown in figure \ref{fig:1}, we  consider convection  in a fluid
layer of thickness $d$, which is inclined 
at an angle $\gamma$ ($0^{\circ} < \gamma <  180^{\circ}$) with respect to
the horizontal.  
Constant temperatures $T_2$ ($T_1$)  with difference $\Delta T = T_1 - T_2
>0$ are prescribed at the boundaries ($z = \pm
d/2)$ of the layer.   Both cases, heating from below ($0^{\circ} <
\gamma \le 90^{\circ}$)   and  heating
from above  ($90^{\circ} < \gamma <  180^{\circ}$), will be considered in this
paper.

The resulting ILC system is  described  by the
standard Oberbeck-Boussinesq equations (OBE) for incompressible fluids. 
As usual, the OBE  are
non-dimensionalized using $d$ as the length scale and the vertical diffusion time 
$t_v= {d^2}/{\kappa}$ as the time scale. 
The velocity $\bm u$ is measured in units of $d/t_{v}$
and  the temperature $T$ in units of $T_s = {\nu
\kappa}/{\alpha  g d^3}$ 
with $\alpha$ the thermal expansion coefficient.
Using a Cartesian coordinate system aligned with the layer (see figure
\ref{fig:1}), the OBE read  as follows:
\begin{subequations}
\label{eq:nondim} 
\beal
 \left[{\partial}/{\partial t}+ (\bm{u} \cdot \bm \nabla) \right] T &= \nabla^2
T  +R
\hat{\bm z} \cdot \bm{u},  \label{eqn:nondimT}\\
\Pran^{-1} \left[{\partial}/{\partial t}+ (\bm{u} \cdot \bm \nabla) \right]\bm u
&=
\nabla^2 \bm{u} - \frac{\bm g}{g}  
T - \bm \nabla p\, , \label{eqn:nondimU}
\end{align} 
\end{subequations}
where $ \bm \nabla \cdot \bm{u} = 0$ due to incompressibility and  
$\bm g = -g \left( \cos\gamma \hat{\bm z}+ \sin\gamma \hat{\bm x} \right)$ describes  
the effect of gravity with the gravitational constant $g$. All terms which can be expressed as
gradients are included in the pressure term   $\bm \nabla p$.
Equations (\ref{eq:nondim})  are characterized by the angle of
inclination $\gamma$ along with two  nondimensional parameters, the
{\it Prandtl} number
$\Pran =\nu/\kappa$ and the {\it Rayleigh} number
$R = \Delta T/T_s$.   

In line with previous theoretical investigations of ILC in the
literature (see in particular  \citep{Clever1977, Busse1992}), we idealize the system to be quasi-infinite in the $x-y$ plane. This is considered to be the appropriate description for large aspect-ratio systems. 
Equations (\ref{eq:nondim}) then admit primary (basic) solutions (denoted with
subscript $0$) of
a linear temperature profile $T_0(z)$ and  cubic shear velocity profile
$\bm U_0(z)$:
\bee
\label{eq:basest}  
T_0(z)=
R \left[\frac{T_1+T_2}{2 \Delta T }-
 z\right], \, \, 
\bm U_0(z) =  \hat{\bm x} \sin \gamma R
\frac{z}{6} \left[z^2 -\frac{1}{4}\right]  \equiv\hat{\bm x } \sin \gamma R
U^x_0(z).
\ee

It is convenient to introduce modifications $\theta$ and $\bm v$ of the basic state and describe the secondary convective state as:
\bee
\label{eq:reduc}
 T(\bm x, z, t) =  T_0(z) + \theta(\bm x, z, t)
\textrm{,}\quad
 \bm{u} (\bm x, z, t) =  \bm{U}_0  + \bm{v}(\bm x, z), \, \, \bm x = (x, y), 
\ee
which fulfill the boundary conditions $\theta(z=\pm 1/2) = \bm v(z = \pm 1/2)
=0$. 
Furthermore, the solenoidal velocity field $\bm v$ is mapped by the well-known poloidal-toroidal decomposition  to two scalar velocity functions $f, \Phi (\bm
x, z, t)$ and a correction $\bm U(z,t)$ of $\bm U_0(z)$; for details, see  
Appendix \ref{App:a}. 
The resulting coupled set of equations for $\theta, f, \Phi, \bm U$ are  analyzed in the
following sections using standard  Galerkin methods and direct
numerical simulations (DNS).

%%%%%%%%%%%%%%%%%%%%%%%%%%%%%%%%%%%%%%%%%%%%%%%%%%%%%%%%%%%%%%%%%%%%%%%%%%%%%%%%%%%%%%%%%%%%%%%%%%%%%%%%%%%%%%%%%%

\section{Finite-amplitude roll solutions}\label{sec:rolls}

Spatially periodic convection roll solutions of the OBE (\ref{eq:nondim}) 
with wavevector  $\bm q$
exist  for Rayleigh numbers
$R > R_c$, where the homogeneous basic state (\ref{eq:basest}) is unstable
against infinitesimal perturbations which depend on $x, y$. The {\it onset of convection}  in ILC system at the  critical Rayleigh number $R_c$, associated with the critical wavevector $\bm q_c$,   has been discussed in \cite{Gershuni1969,birikh1972, Hart1971}. A very useful overview  can be found in   \cite{Chen1989} and references therein. Some additional general information is given in  Appendix \ref{App:clin}.

Since the ILC system is
anisotropic, we have to consider the  linear stability of the basic state
against arbitrarily-oriented convection rolls with wavenumbers $\bm q = q(\cos
\psi, \sin \psi)$. For that purpose, we have to analyse (\ref{eq:nondim})
linearized about the basic state (\ref{eq:basest}). For details of the
standard numerical method, see Appendix
\ref{App:alin}. For $\gamma =0$ (horizontal layer with $\bm U_0 \equiv 0$)
the system is isotropic and we have the standard
Rayleigh-B{\'e}nard convection (RBC) where $|\bm q_c|  = q_{c0} = 3.1163$ and
$R_c = R_{c0} = 1707.762$ (see e.g. \cite{Ahl1984}) which depend on neither $\psi$ nor $\Pran$.   
This is distinct from finite $\gamma$,   since  $\bm U_0(z)$ defined in (\ref{eq:basest}) yields a
contribution  proportional to
$\cos \psi \sin \gamma/\Pran$ in the linear equations (see \ref{eq:incfn}).  

In the following, we concentrate on the special case $\Pran =1.07$, where
the bifurcation of the basic
state is always stationary; other $\Pran$ are briefly discussed In Appendix \ref{App:cobl}. 
Figure \ref{fig:2} displays the rescaled critical Rayleigh number
$R_c /R_{c0}$ 
and the critical wavenumber $q_c$ as function of the inclination angle
$\gamma$ 
and different $\psi$.
In general, only two particular
$\bm q-$orientations turn out to be relevant (see e.g. Appendix \ref{App:cobl}).
The convection
solutions at onset are either
buoyancy driven {\it
longitudinal} rolls with their axes along the incline, i.e. $\bm q_c = q_c
\hat{\bm y},~\psi = 90^{\circ}$  or shear driven {\it transverse} rolls with
their axis perpendicular to the incline where $\bm q_c = q_c  \hat{\bm x},~\psi
= 0^{\circ}$.

Longitudinal rolls ($\psi = 90^{\circ}$) exist only 
in the range $0< \gamma < 90^{\circ}$ (heating from below, see figure
\ref{fig:1}). Their critical wavenumber is given by $q^l_c = q_{c0}$ for all
$\gamma$  and the critical Rayleigh number $R_c^l(\gamma)$ fulfils the relation
$R^l_c(\gamma)\cos\gamma = R_{c0}$ (see Appendix \ref{App:cobl}), implying
that $R^l_c$ diverges in the limit $\gamma \rightarrow 90^{\circ}$.
In contrast, a bifurcation to transverse rolls  exists in the whole interval
$0 < \gamma < 180^{\circ}$.  The critical Rayleigh number $R^t_c(\gamma)$
rises continuously as function of $\gamma$ and diverges at $\gamma =
180^{\circ}$ (stable horizontal fluid layer, heated from above).
In figure \ref{fig:2}  the critical data have
been shown only for $\gamma$ up to $120^{\circ}$, where 
$R^t_c\sim 10^5$ involves large thermal gradients.  Thus, the use
of the OBE becomes questionable for $\gamma >120^{\circ}$, since 
non-Boussinesq effects due to temperature variation of the various material
parameters should be taken into account.  

Inspection of figure \ref{fig:2}  reveals the existence of  a codimension-2
bifurcation point $\gamma_{c2} = 77.746^{\circ}$ where $R_c^l = R^t_c =8046.420$,
such that for $\gamma < \gamma_{c2}$  longitudinal rolls bifurcate at onset 
( $R_c^l < R^t_c$) while for  $\gamma > \gamma_{c2}$ the transverse ones
prevail. As first demonstrated in \cite{Gershuni1969} and detailed in Appendix
\ref{App:cobl}, the threshold curves $R^{ob}_c(\gamma, \psi)$ for general oblique rolls ($\psi \ne 90$)  can be constructed  by suitable transformations of the critical values $R^t_c(\gamma)$ and  $q^t_c(\gamma)$ of the transverse rolls. In this paper, we will often use the reduced main control parameter $\epsilon$ defined as:
\begin{equation}
\label{eq:defeps}
\epsilon =(R -R_c(\gamma, \Pran)) / R_c(\gamma, \Pran)  
\end{equation}
as a measure for the relative distance from threshold $R_c(\gamma)$ at $\epsilon
=0$, instead of $R$.

%%%% figure 2
\begin{figure}
\vspace{0.4cm}
\centering{\includegraphics[width=12cm]{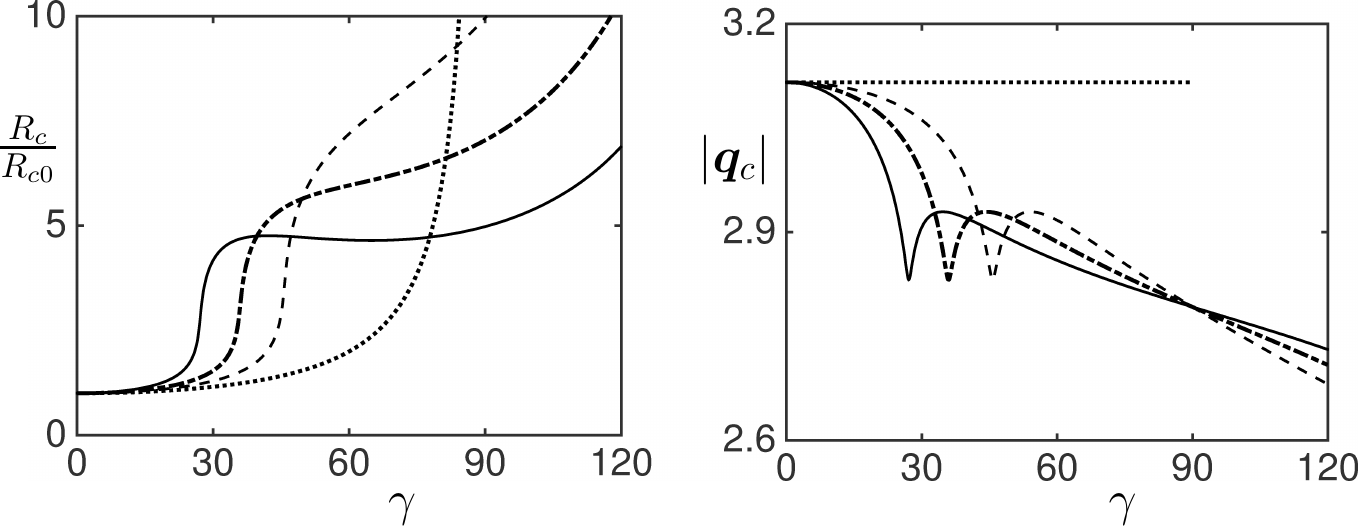}}
\caption{Normalized critical Rayleigh number
$R_c/R_{c0}$ (left panel) and critical wave number $|\bm q_c|$ (right panel)
as function of the 
inclination angle $\gamma$ for $\Pran = 1.07$ for different roll orientations 
$\psi$. Dotted line is for $\psi=90^{\circ}$ (longitudinal rolls),
dashed line for $\psi=60^{\circ}$, dashed-dotted line for
$\psi=45^{\circ}$ and solid line for $\psi=0^{\circ}$ (transverse rolls).}
\label{fig:2}
\end{figure}

The standard computational methods  to construct  finite-amplitude roll
solutions with wavevector $\bm q_c$ for $R > R_c$, where  exponential growth of
the linear modes is balanced by the nonlinear terms in the OBE, are sketched in Appendix \ref{App:asecond}.
The amplitudes of the roll solutions grow continuously like
$\sqrt{\epsilon}$ (see  the discussion after (\ref{eq:amp})).
Thus, the primary bifurcation to rolls is continuous (forward).

%%%%%%%%%%%%%%%%%%%

%\input{section4_mar3.inp.tex}
\section{Secondary instabilities of roll solutions for
\Pran = 1.07}\label{sec:secstab}
\begin{figure}
\begin{center}
{\includegraphics[width=12cm]{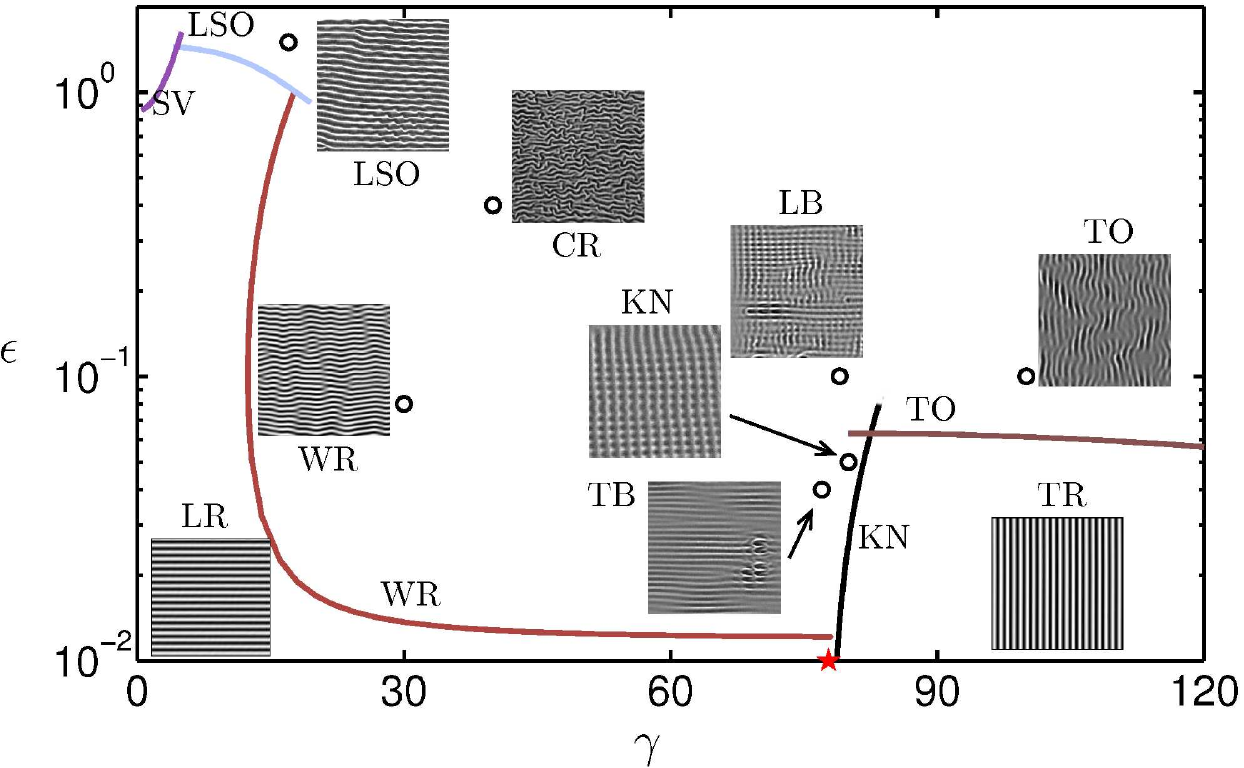}}
\end{center}
\caption{Phase diagram of the convective roll patterns in the
$\gamma, \epsilon$ plane for  $\Pran=1.07$.
Solid lines (coloured online) indicate the  secondary instabilities of the primary roll
patterns  at $\epsilon = \epsilon_{inst}$, i.e. at $R = (1 + \epsilon_{inst}) R_c(\gamma)$. The codimension 2 point with $\gamma_{c2} =77.746^\circ$ is marked by a star. For $\gamma<\gamma_{c2}$, we obtain longitudinal rolls (LR), and for $\gamma > \gamma_{c2}$, transverse rolls (TR). Increasing $\gamma$ in the interval $0 \le \gamma <  \gamma_{c2}$, we obtain the thresholds for skewed varicose instability (SV) shown in purple, longitudinal subharmonic oscillations (LSO) shown in light blue and wavy rolls (WR) shown in dark red as instabilities of the LR. In the range $\gamma > \gamma_{c2}$ the thresholds for instability of the transverse rolls through the knot instability (KN) is shown in black and for the transverse oscillations (TO) in brown. The graph is decorated with excerpts from 
corresponding  experimental pictures \citep{Karen2000}. In addition crawling rolls (CR), transverse (TB) and the longitudinal bursts (LB) are shown, which cannot be directly associated to the secondary instabilities.   
Black open circles indicate locations in the
$\gamma-\epsilon$ plane, where experiments and numerical simulations are compared in section \S{\ref{sec:comp}.}}\label{fig:3}
\end{figure}

%%%%%%%%%%%%%%%%%%%%%%%{\color{red} Introduce nonlinear patterns}\\
In this section, we discuss the secondary destabilisation mechanism of rolls with wavevector $\bm q_c$, in a  ILC  system with $\Pran = 1.07$ and inclination angle $\gamma$,  that become  unstable when $\epsilon = \epsilon_{inst}(\gamma)$. Based on the methods described in Appendix \ref{App:asecond}, the stability diagram presented in figure \ref{fig:3}  has been determined in the $\gamma-\epsilon$ plane.The solid lines mark the locations of the various secondary 
instabilities  of the finite amplitude roll solutions with $\bm q = \bm
q_c(\gamma)$ at $\epsilon = \epsilon_{inst}(\gamma)$. Thus to the left of this line for $\gamma \approx 15^{\circ}$  and below this line for larger $\gamma$ stable roll solutions exist. For details of the dependence of these secondary stability lines  upon cut-off parameters in the Fourier-Galerkin expansions see Appendix \ref{App:astabaccur}.
 
According to figure \ref{fig:3}, the type of secondary roll instabilities depends strongly on the inclination angle $\gamma$. Our main interest  in this paper are the various  $3D$ patterns which develop for $\epsilon >\epsilon_{inst}$. As a first impression, we include in figure \ref{fig:3} excerpts of patterns observed in experiments \citep{Karen2000}. Here, we aim to reproduce and interpret such $3D$ patterns based on direct numerical simulations (DNS) of the OBE. For details of time stepping scheme we refer to the algorithm presented in Appendix \ref{App:B}.

\begin{figure}
\centering{
{\includegraphics[width = 10cm]{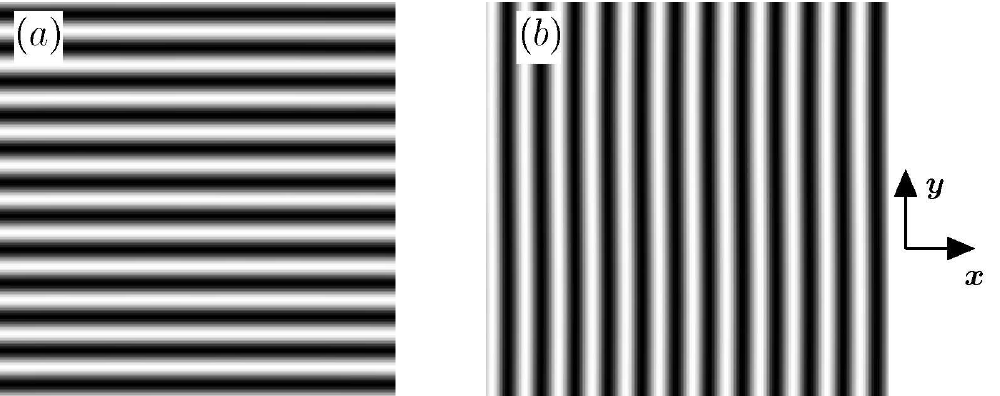}}
}
\caption{Temperature plots $\langle\theta(\bm x)\rangle$, for $\Pran = 1.07$   from  direct numerical simulations of the OBE (\ref{eq:nondim}): ($a$) buoyancy dominated longitudinal rolls oriented along the incline from
right to left at $\Pran=1.07$, $\epsilon=0.01$, $\gamma=10^{\circ}$ with  $\bm {q_c}=(0,3.1163)$ (left panel); ($b$) shear dominated transverse rolls oriented perpendicular to the incline at $\epsilon=0.02$, $\gamma=85^{\circ}$ with $\bm{q_c}=(2.82,0)$ (right panel). In the plane the figures are oriented along the x-axis of  local coordinate system in figure \ref{fig:1} (reproduced at the right). All temperature plots in the rest of the paper are shown using the same convention.}
\label{fig:4}
\end{figure}

We first consider small $\epsilon <\epsilon_{inst}(\gamma)$. 
Starting from random initial conditions, modes with wave
vector $\bm q_c(\gamma)$ prevail leading to perfect roll patterns, as shown in figure \ref{fig:4}.
The DNS are performed on a square with side lengths $L_x = L_y = 12 \lambda_c$ with $\lambda_c = 2 \pi/qc(\gamma)$ where we obtain, longitudinal rolls at $\gamma = 10^{\circ}$ and transverse rolls at $\gamma
= 85^{\circ}$. Here, and in the rest of this paper, we show snapshots of the
vertical ($z$) average $\langle\theta(\bm x,t)\rangle$ of the  temperature field $\theta
(\bm x, z,t)$ (see (\ref{eq:galerk})). Throughout this paper the height of the convection cell increases from left to right, i.e. with increasing $x$ with respect to the coordinate system attached to the cell (see figure \ref{fig:1}). Such pictures, which we refer to as the temperature
plots, are  typically used in the literature  to compare with
experimental convection patterns, which are visualized via shadowgraphy (for
examples, see  \cite{Bodenschatz2000}). In gas convection experiments of the type considered in this
paper, dark and bright regions in figure \ref{fig:4} indicate positive (hot) and
negative (cold) variations in the temperature field around the basic linear
temperature profile \citep{deBruyn1996}. We use a 8-bit grayscale to
visualize  $\langle\theta(\bm x)\rangle$; whose range increases  monotonically as a function of $\epsilon$. 

%%%%%%%%%%%%%%%%%%%%%%%%%%%%%%%%%%%%%%%%%%%%%%%%%%%%%%%%%%%%%%%%%%%%%%%%%%%%%%%%
%%%%%%%%%%%%%%%%

%%%%%% figure 3

%%%%% figure 4

In the following, we discuss the secondary instabilities of
the primary convection rolls in detail. We start with inclination angles $\gamma$ below $\gamma_{c2}$ in 
\S\ref{sec:secbelow}, before  we concentrate on the vicinity of $\gamma_{c2}$
and finally on larger $\gamma$.  
%For convenience we use the representation $R_{inst} = R_c(\gamma) (1 + \epsilon_{inst})$ (see (\ref{eq:defeps})). 

The various secondary instabilities are visualized by direct simulations of
the underlying OBE (see Appendix \ref{App:btime}). In general  we use a 
{\it minimal} rectangular integration domain  in the $x-y$ plane which is
consistent
with $\bm q_c$ and the wavevectors of the dominant destabilizing modes. For 
visualisation, the domain is periodically extended to a larger domain  
with $Lx = 12 \lambda_c = L_y$.

%%%%%%%%%%%%%%%%%%%%%%%%%%%%%%%%%%%%%%%%%5

\subsection{Secondary roll instabilities below $\gamma_{c2}$}
\label{sec:secbelow} 
In this section, we will characterize in detail the secondary instabilities of
longitudinal rolls in figure \ref{fig:3}.

\subsubsection{Skewed varicose instability  (SV)}\label{sec:subsv}
For small inclinations ($\gamma \lesssim  5^{\circ}$), we recover the well known skewed varicose (SV) instability for planar RBC with
$\gamma =0$ \citep{Busse1979}. This is a stationary long-wavelength instability 
where the original longitudinal rolls are slowly modulated along their axes
but also with respect to their distance. The SV instability will not be further discussed in
this paper. 

\subsubsection{Logitudinal subharmonic oscillations (LSO)}\label{sec:subhar}
%%%%%%%%%%%%%%%%{\color{red} Short wave instability - Subharmonic Busse oscillations}\\

%%%%%%% figure 5
\begin{figure}
\centering{
{\includegraphics[width = 4.1cm]{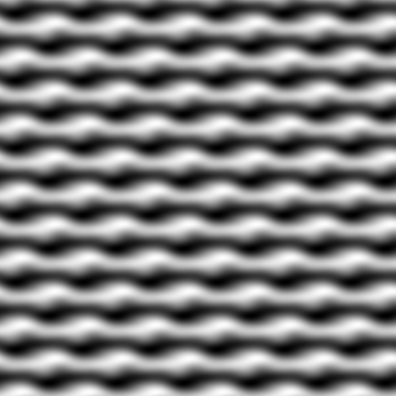}}
\hspace{1cm}
{\includegraphics[width = 4.1cm]{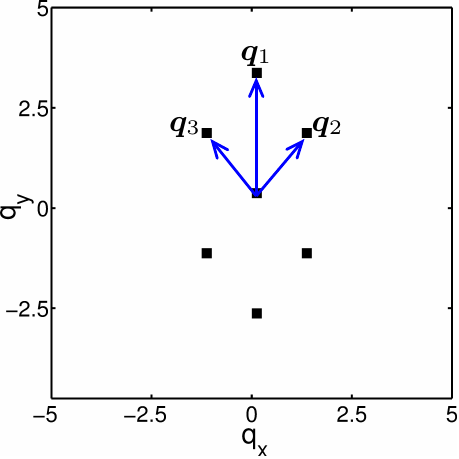}}
}
\caption{Subharmonic oscillatory instability of longitudinal rolls (LSO) for
$Pr=1.07$, $\gamma=17^{\circ}$ and $\epsilon=1.3$: temperature plot (left panel) together with the wavevectors $\bm q_1 =\bm q_c, \bm q_2, \bm q_3$  (wavevector resonance $\bm q_1 = \bm q_2 + \bm
q_3$)  of the  leading Fourier modes (right panel). }.
%For $q_x=1.28$ and $q_y=q_c=3.117$ and for a chosen number of rolls $n_L=12$, we have $\triangle q_x=q_x/n_L=0.2134$ and $\triangle q_y=q_y/n_L=0.2598$. }
\label{fig:busse}
\end{figure}

In the range $5^{\circ}<\gamma\leq 21^{\circ}$, the longitudinal roll pattern
with $\bm q =  \bm q_c = (0,3.1163)$ becomes linearly unstable to
oscillatory subharmonic perturbations with wavevectors  $\bm q_{2,3} =(\pm q_x,
q_c/2)$ and a finite circular frequency $\omega_{inst}$. For the representative
case $\gamma =
17^{\circ}$, where primary rolls get unstable at $\epsilon_{inst} = 1.044$, we
have
$q_x = 1.279$ and $\omega_{inst} = 10.21$. In figure \ref{fig:busse}, we show an excerpt from our simulation at $\epsilon = 1.3$.
performed on the minimal  rectangle with sidelengths $L_x = 2  \pi /q_x$ and
$L_x =
2 \lambda_c /q_c$ with $\lambda_c = 2 \pi/q_c$. The periodically extended picture is six times larger.
The pattern is characterized
by periodic modulations  of the longitudinal rolls, which are in phase on
every 
second roll  reflecting  the subharmonic nature of the instability. The resulting LSO pattern obeys the wavevector resonance $\bm q_1  -\bm q_2 - \bm q_3 =0$
as seen in the  Fourier spectrum.  
%The minimal integration domain, as  described before,  had in this case  the extension $L_x = 2 \pi /{q_x}, L_y = 4 \pi /q_c$. 

\begin{figure}
\hspace{-0.5cm}\centering{\includegraphics[width=9cm]{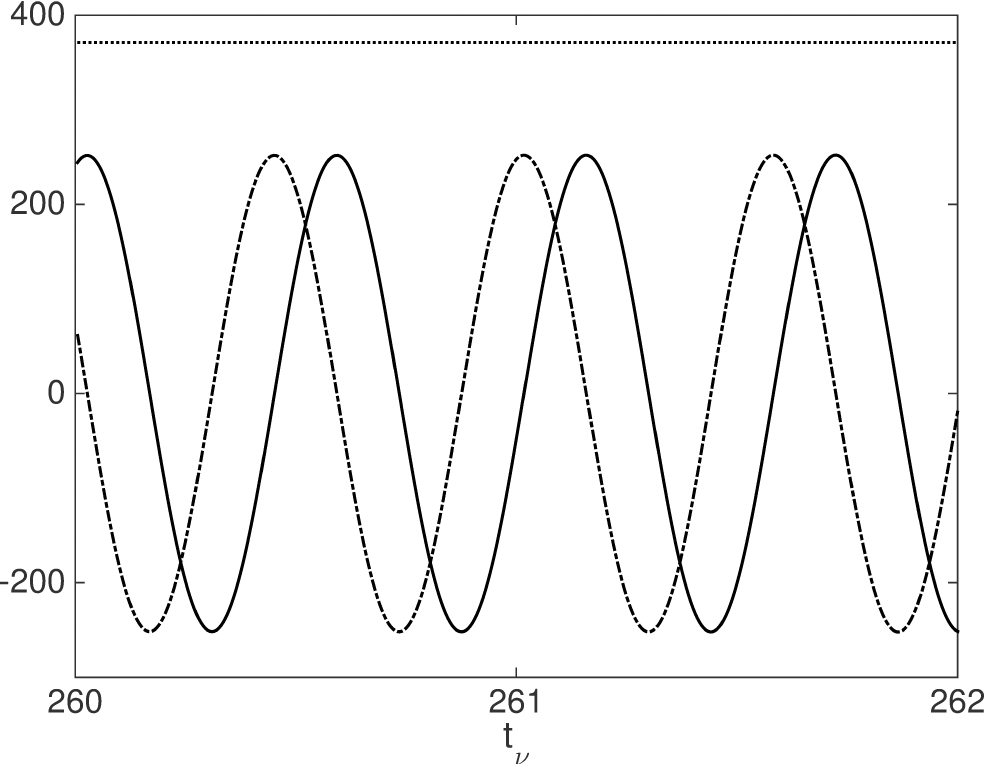}}
\caption{The coefficients $A$ (horizontal upper dotted line),
$B(t)$
(dash-dotted) and $C(t)$ (solid line) in (\ref{eq:patsv}) as function of
time in units of $T_s$;   $B(t), C(t)$ are multiplied by a
factor of four for better visualisation.}
\label{fig:shvABC}
\end{figure}

The time evolution of the pattern in the $x-y$plane is:
\begin{equation}
f(x,y) = A \cos(q_c y) + B(t) \cos(q_x  x + q_c y/2  ) + C(t)  \sin(q_x  x - q_c
y/2 )
\label{eq:patsv}
\end{equation}
The coefficient $A$ is independent of time while $B(t)$ and $C(t)$ are periodic 
with circular frequency $\omega \approx \omega_{inst}$. The time evolution of these
coefficients (given in units of $T_s = O(1m K)$ defined in \S{\ref{sec:gde}}) are shown in figure \ref{fig:shvABC}.

%%%%%%%%%% figure 6

The LSO  instability has
been first
described by \cite{Busse2000} 
at  the lower Prandtl numbers $\Pran=0.7$, by solving
(\ref{eq:four2d}) in the  minimal integration domain.   In addition for certain parameter combinations of $q_x, q_y$ an
intermittent appearance of
bursts has been described, which we have not reproduced for $\Pran=1.07$.
% in our case 

%%%%%%%%%%%%%%%%%%%%%%%%%%%%%%

\subsubsection{Wavy Roll (WR) instabilities}\label{sec:wavy}
%%%%%%%%%%%%%%%%%%%%%%%{\color{red} Wavy instabilities }\\
The next instability type, characterized by the appearance of 
longitudinal rolls undulating like snakes along their axes,   has been first
described in \cite{Clever1977} where the
notion wavy-instability has been coined.
The resulting bifurcation to  wavy rolls (WR) is observed in a fairly large
$\gamma$-interval between
$21^{\circ} < \gamma \leq \gamma_{c2}$, very close to onset of convection (
$\epsilon_{inst} = O(0.01)$).   In the framework of the Galerkin stability
analysis in Appendix \ref{App:asecond}, this instability is characterized by long-wavelength
destabilizing modes with wavevectors $\bm q_{max} =(\pm q_x, q_c)$ with 
($|q_x| \ll |\bm q_c|$).  

The WR have been discussed in detail in  \cite{Karen2008}, to which we refer
for details. Here one finds
representative experimental and theoretical pictures (see also figure
\ref{fig:ucT} in \S\ref{sec:comp} below) as well as a Galerkin stability analysis
of the 3D wavy roll patterns. In fact,   for  $\epsilon > \epsilon_{inst}$  
 stable WR  with finite $q_x$ exist. They  are
spatially periodic in the plane,  
characterized by the   wavevector resonance $\bm q_2 + \bm q_3
= 2 \bm q_c$  with $\bm q_{2,3} = (\pm q_x, q_c)$.
In the $x,y-$plane the temperature pattern is 
described as 
\begin{equation}
f(x,y) = A\cos( q_c y) + B\sin ( q_c y)\sin(q_x x ).
\label{eq:patwavy}
\end{equation}
For $\epsilon = 0.1$ and $q_x = 1.28$, we find for instance $A
=48.1$ and $B=12.8$  in units of $T_s$.

%%%%%%%%% figure 7
\begin{figure}
\centering{
{\includegraphics[width = 4.1cm]{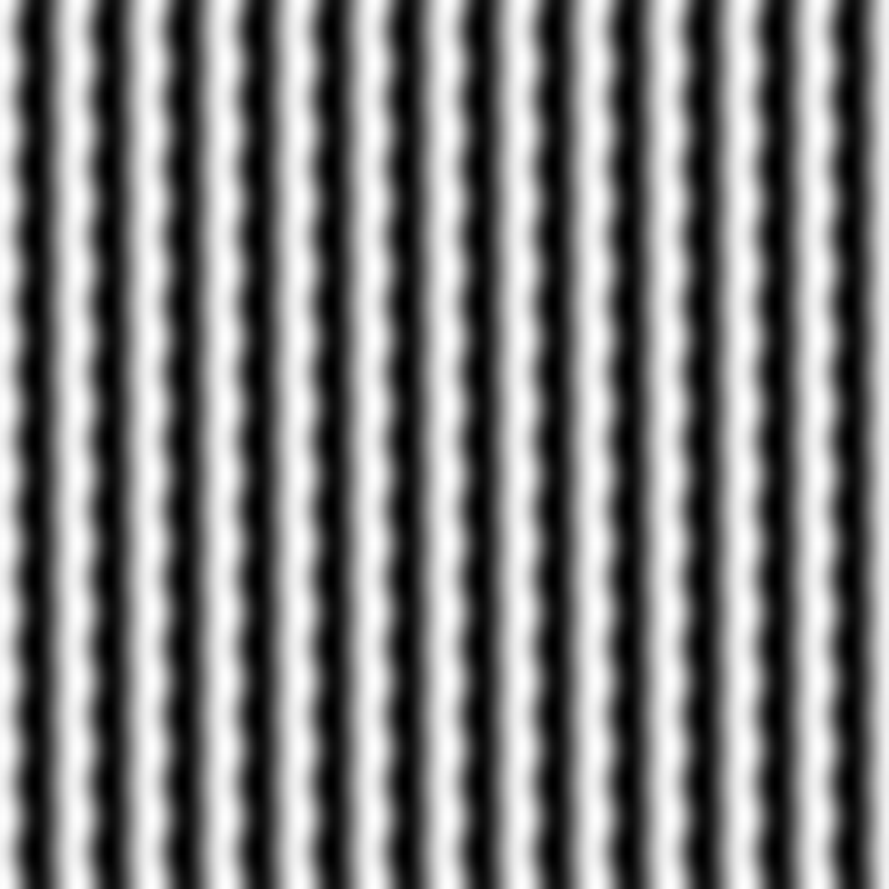}}
\hspace{1cm}
{\includegraphics[width = 4.1cm]{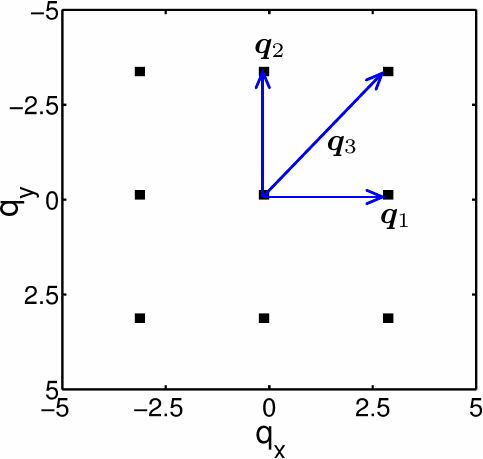}}
}
\caption{Knot instability of transverse rolls  at $Pr=1.07$,
$\gamma=81.9^{\circ}$ and $\epsilon=0.055$: 
temperature plot (left panel);  wavevectors of the leading Fourier modes $\bm
q_1 = \bm q_c$, $\bm q_{2} =
(0, q_{c0})$  and ${\bf{q}}_3\,=\,{\bf{q}}_1+{\bf{q}}_2$ (right panel).}
%Given that $q_x=2.82$ and $q_y=q_c=3.117$ and choosing the number of rolls $n_L=12$,
%we have $\triangle q_x=q_x/n_L=0.235$ and $\triangle q_y=q_y/n_L=0.2598$.}
\label{fig:knot}
\end{figure}

\subsection{Secondary roll instabilities above $\gamma_{c2}$}
In this section we discuss the secondary instabilities of the transverse rolls
bifurcating for inclinations $\gamma > \gamma_{c2}$.

\subsubsection{Knot (KN) instability}\label{sec:knot}
%%%%%%%%%%%%%%%%%%{\color{red} Cross roll instability}\\
Just above the codimension 2 point, the shear dominated
transverse rolls with $\bm{q_1}=(q_c,0)$ and $q_c \simeq 2.82$ (see figure
\ref{fig:2}) are destabilized by
the longitudinal rolls with 
wavevector $\bm q_2= (0,q_y)$ and $q_y= q_{c0}= 3.1163$.  The steeply rising
stability line starts at $\gamma = \gamma_{c2}$ at $\epsilon =0$. In view of the
logarithmic $\epsilon$-scale, the curve is only shown for $\epsilon >0.01$. In the weakly
nonlinear regime, the oblique mode with wave vector $\bm q_3= (q_c,q_y)$
is already important and the wavevector resonance 
$\bm q_3\,=\,\bm q_1+\bm q_2$ is established. 
The pattern is well described by:
\begin{equation}
f(x,y) = A\cos( q_c x) + B\sin (q_y y) +  C \sin (q_c x)\cos(q_y y)
\label{eq:patknot}
\end{equation}
As a function of $\epsilon$, the amplitudes $B, C$ increase 
continuously above $\epsilon = \epsilon_{inst}$. 

In figure \ref{fig:knot} we address an representative example for $\gamma =
81.9^{\circ}$ with $\epsilon_{inst} = 0.053$, 
where  originally $Lx = \lambda_c$ and $L_y =
2\pi/q_{c0}$.  
At  $\epsilon = 0.055$,  we find $A = 17.9, B = 1.95, C = 2.16$ in
(\ref{eq:patknot}). The resulting stationary pattern (see figure  
\ref{fig:knot}, left panel) has some similarity to the knot patterns described
in \citet{Busse1979} for  the isotropic RBC system. However, contrary to the patterns in 
\citet{Busse1979}, we observe a wavevector resonance triggered by the oblique mode $\bm{q}_3$ in the ILC system. 

For completeness, it should be mentioned that the knot instability in ILC has
been previously investigated in
\citet{Fujimura1992} for $\gamma \lesssim 90^{\circ}$ in the framework
of two coupled amplitude equations restriced  to  the amplitudes $A, B$ in 
(\ref{eq:patknot}).

\subsubsection{Transverse Oscillatory rolls (TO)}\label{sec:sdp}
\begin{figure}
\centering{
{\includegraphics[width = 4.1cm]{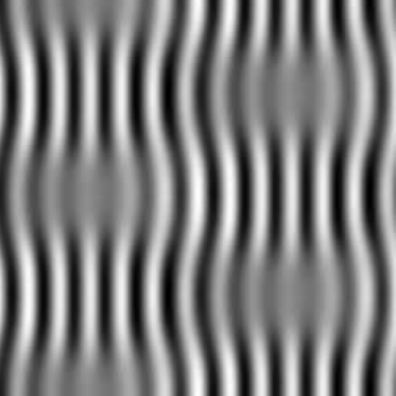}}
\hspace{1cm}
{\includegraphics[width = 4.65cm]{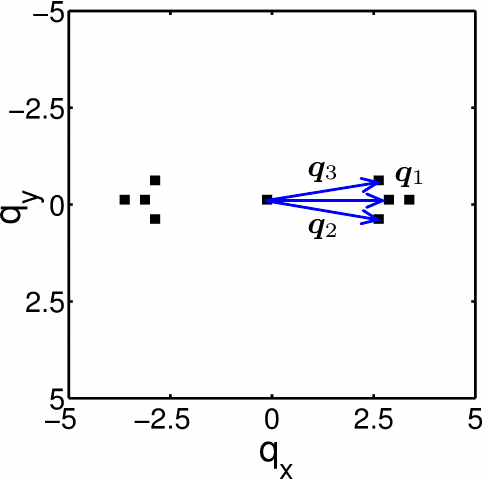}}
}
\caption{Transverse oscillatory rolls at $Pr=1.07$,
$\gamma=84.9^{\circ}$ and $\epsilon=0.07$: temperature plot (left panel); wave vectors of the dominant   Fourier
modes (right panel). In this case, we use a square grid with $q_y=q_c=3.117$. 
%Choosing the number of rolls $n_L=12$, we have $\triangle q_x=\triangle q_y=q_y/n_L=0.2598$.
}
\label{fig:sdp}
\end{figure}

For $\gamma>83.2^{\circ}$ the destabilization of the transverse rolls  starts to be governed  by the
transverse oscillatory rolls (TO)  along an  almost horizontal
transition line  as function of $\gamma$. Transverse oscillatory rolls (TO) are
characterized by  destabilizing modes with a Floquet vector $\bm s$ of
relatively small but finite modulus  $|\bm s|  \sim q_c/6$ and by an
oscillatory time dependence  of period  about $3.5 t_{v}$.   
In an analogy to the stationary SV instability 
of longitudinal rolls (see \S\ref{sec:subsv}), the rolls are expected to
become slowly modulated along their axis and also with respect to other in-plane 
direction.
Without a definitive resonance condition among the dominant destabilizing modes,
we do not expect periodic  3D patterns of the kind 
discussed in the previous subsections. Thus, we have to solve the OBE on a larger domain in the
$x-y$ plane,    including modes with wave
vectors $|(q_x, q_y)| \ll q_c$.  
An excerpt of representative DNS pattern for $\gamma = 84.9^{\circ}$ is shown in figure \ref{fig:sdp} (left panel) at $\epsilon=0.07$. This pattern results from a secondary instability of transverse rolls ($R_c = 8282.64$ and $q_c = 2.8023$) at $\epsilon_{inst} = 0.063$ with $\omega_{inst} = 1.809$. One observes the 
appearance of
localized patches with reduced amplitudes  on top of the original
slowly modulated transverse rolls. Apparently, the pattern arises through a complex beating phenomena, due to a superposition of oscillating modes with slightly different wave vectors. The most prominent ones ($\bm q_2, \bm q_3$) together with $\bm q_1 \equiv \bm
q_c$ are shown in figure \ref{fig:sdp} (right panel). 

\begin{figure}
\centering{
{\includegraphics[width = 9cm]{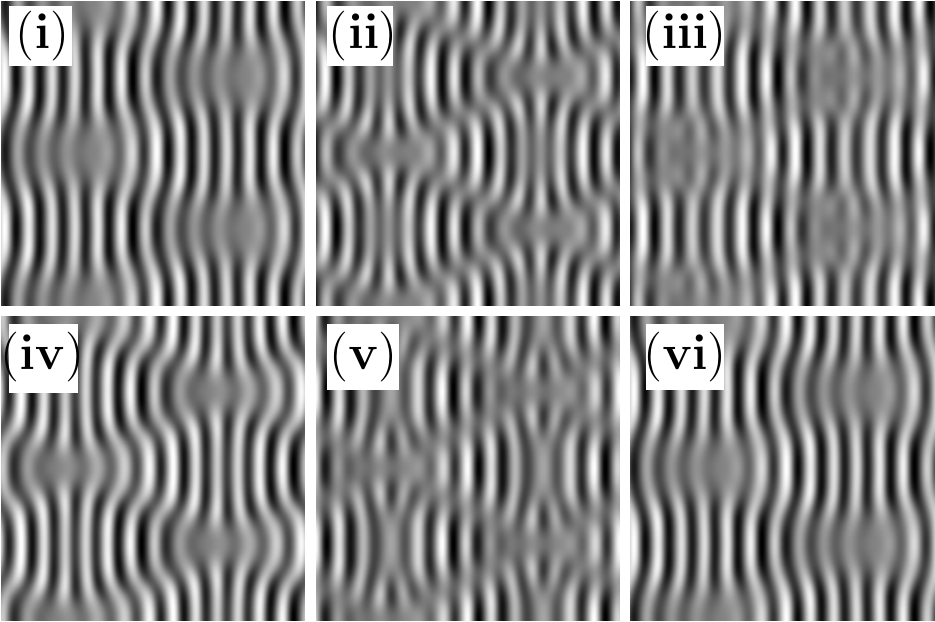}}}
\caption{A time sequence of TO patterns shown over one time period for
the case of heating from below. Consecutive panels are separated by $0.44
t_{\nu}$. System parameters are the same as figure \ref{fig:sdp} with $Pr=1.07$,
$\gamma=84.9^{\circ}$ and $\epsilon=0.07$.}
\label{fig:sdptime1}
\end{figure}

Figure \ref{fig:sdptime1} shows the complicated evolution of the pattern during one cycle ($2 \pi/\omega_{inst}$). The localized patches of reduced amplitudes in figure \ref{fig:sdptime1}(i), evolve into slanted lines of reduces amplitudes in (ii). Further in the cycle, the undulation amplitude of the rolls first increases and decreases then before arriving again at the  initial pattern in figure \ref{fig:sdptime1}(iii-vi).

\begin{figure}
\centering{
{\includegraphics[width = 13cm]{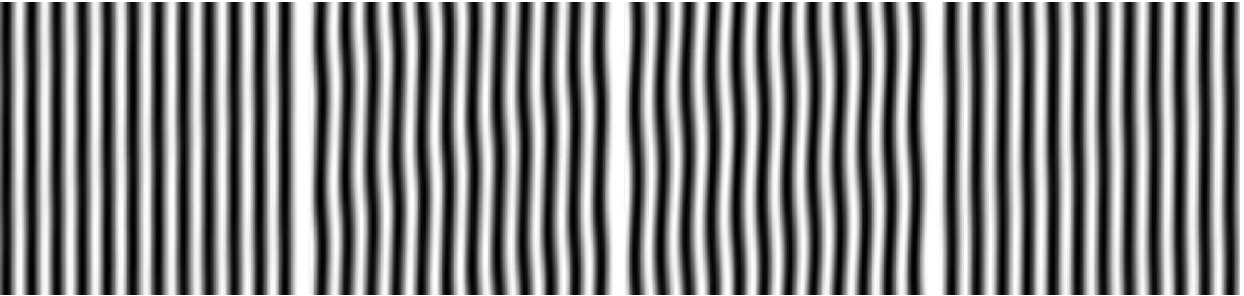}}}
%{\includegraphics[width = 13cm]{./figs/fig11.eps}}}
\caption{ A time sequence of TO patterns shown over one time period for
the case of heating from above. Consecutive panels are separated by
$0.4 t_{\nu}$. System parameters are $Pr=1.07$, $\gamma=100^{\circ}$ and
$\epsilon=0.08$.}
\label{fig:sdptime2}
\end{figure}

According to our phase diagram in figure \ref{fig:3}, the instability of the transverse rolls towards the TO pattern remains relevant for larger $\gamma$ and also governs the secondary instability in the case of heating from above. A representative example of a time sequence is shown in figure \ref{fig:sdptime2} for $\gamma=100^{\circ}$, where the TO instability is characterized by  $\epsilon_{inst} = 0.06$ and $\omega_{inst} =1.48$.  The graph of the 
most relevant destabilizing modes looks practically identical to the one in
figure  \ref{fig:sdp} and is thus not shown. The interaction of the modes leads,
however, to a much simpler time evolution, as compared to  figure
\ref{fig:sdptime1} for $\gamma = 84.9^{\circ}$.  It is possible that in the
latter case, the destabilizing modes triggering the knot instability for slightly
smaller $\gamma$ in figure \ref{fig:3} come into play as well.

As will be discussed in  \S{\ref{sec:comp}}, the regions of suppressed
amplitudes become elongated and  are no longer periodically arranged in the
plane for increasing aspect ratio ($L_x, L_y$).  As shown in figure
\ref{fig:sdpT}  below, they compare well with
the corresponding experimental patterns in  \citep{Karen2000}, called 
switching diamond panes (SDP) there.
  
%%%%%%%% figure 8

%%%%%%% figure 9

%%%%%%% figure 10

\subsubsection{Vertical convection}\label{sec:vert}
%%%%%%%%%%%{\color{red} vertical case}\\
The case of a vertical convection cell ($\gamma=90^{\circ}$) is of special
interest since the pattern formation  is exclusively driven by the shear
stress. So this system has
motivated many previous investigations, mainly in the linear regime
and often with
$\Pran \gtrsim 12.45$ where
an oscillatory bifurcation to transverse rolls takes place.
For $\Pran = 1.07$, however,  we observe stationary transverse rolls  at onset,  and
their 
stability analysis  yields always a secondary  
bifurcation to the TO pattern for all $\gamma$ near  $90^{\circ}$ (see figure
\ref{fig:3}  and our discussion in the previous subsection).   

This result is noteworthy, since in the  previous literature \citep{Busse1995} for
$\Pran = 0.71$ (air)  a  stationary secondary instability of the transverse
rolls driven  by the 
effective subharmonic roll modes with wavevectors $\bm q_{2,3} =(q_c/2, \pm p)$ was predicted. This finding   has been confirmed
by our own calculations.  The instability of the  transverse rolls with 
$q_c =  2.8123, R_c = 5701.2625$ takes place at $\epsilon_{inst} = 0.0599$ with
$p = 1.5898$. These numbers are consistent with those used  for direct
simulations of the OBE  in \citet{Busse1995} ($R_c = 5726.9, q_c = 2.69, p =
1.7$ and $\epsilon \gtrsim 0.11$).   
 
\begin{figure}
\centering{
{\includegraphics[width = 4.1cm]{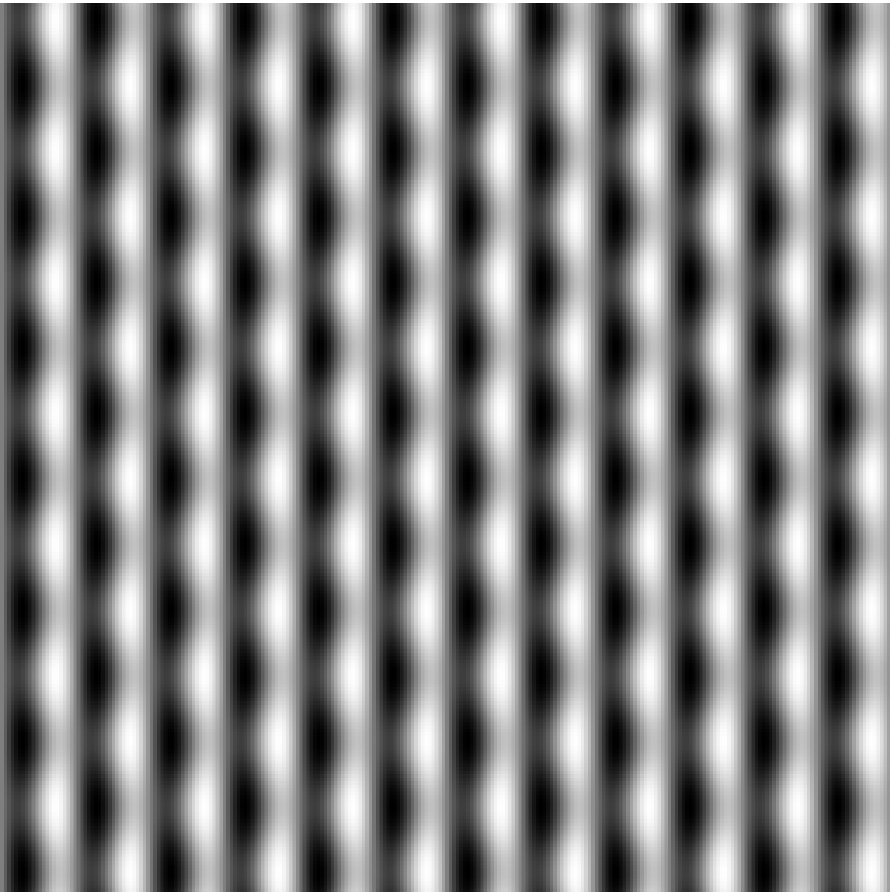}}
\hspace{1cm}
{\includegraphics[width = 4.1cm]{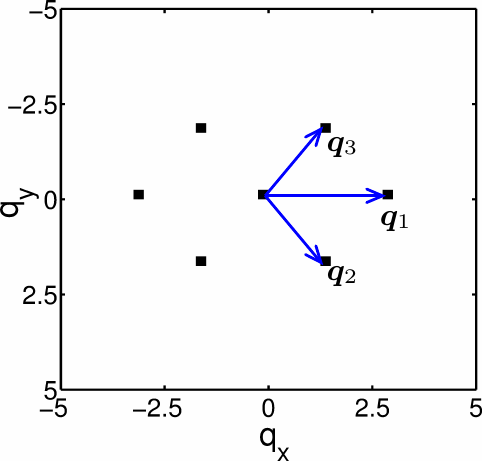}}
}
\caption{Subharmonic varicose instability of transverse rolls with $\Pran = 0.71$, $\epsilon = 0.065$ and $\gamma = 90^{\circ}$:
temperature plot (left panel); wavevectors of the leading
Fourier amplitudes (right panel). Given that $q_x=2.8123$ and $q_y=1.59$ and choosing the number of rolls $n_L=12$, we have $\triangle q_x=q_x/n_L=0.2344$ and $\triangle q_y=q_y/n_L=0.1358$.}
\label{fig:shv}
\end{figure}

 In close analogy to the LSO (\S\ref{sec:subhar}), the
resonance conditions $ \bm q_c =  \bm q_{2} + \bm
q_{3}$ holds.  To confirm the results of stability analysis we have performed
DNS of the OBE on the smallest periodicity domain in the 
plane compatible with instability data above, i.e. with  $L_x = 2 \lambda_c,
L_y= \lambda_c (q_c/p)$. The resulting  stationary temperature plot
(periodically extended) for $\epsilon = 0.064 > \epsilon_{inst}$ is shown in 
figure \ref{fig:shv} for $L_x = L_y = 12 \lambda_c$.
Keeping the dominant Fourier modes, the pattern is represented as:
\begin{equation}
f(x,y) = A\cos( q_c x) - B \sin (q_c x/2 )\cos(p  y )
\label{eq:patshv}
\end{equation}
with $A = 131.23, B =60.76$ in units of $T_s$; the subharmonic 
instability of the transverse rolls is obvious from the argument of sine in the second term. 

A closer look at the $\Pran$-dependence of the secondary bifuraction of the
transverse rolls 
in this regime shows that for $\Pran
\gtrsim 0.9$ the secondary SHV bifurcation  of the transverse rolls is replaced
by the TO bifurcation discussed in 
\S\ref{sec:sdp}. This is consistent  with our stability analysis and experimental observations at $\Pran = 1.07$.

\vspace{0.5cm}

 %%%%%% figure 11
%%%%%%%%%%%%%%%%%%%%%%%%%%%%%%%%%%%%%%%%%%%%%%%%%%%%%%%%%%%%%%%%%%%%%%%%%%

\section{Comparison with experimental results}\label{sec:comp}

In the previous section, we  have discussed  the various characteristic
secondary instabilities of the ILC roll patterns with increasing inclination
angle $\gamma$.  A number of basic destabilization mechanisms 
have been identified  by considering simulations
in small periodicity domains in the plane of linear dimension $L$, 
where $L =O(2 d)$ with $d$ the cell thickness.
Our goal in this section is a comparison with the pressurized $\mathrm{CO_2}$
experiments \citep{Karen2000,karenthesis2002} at $\Pran =1.07$. In these experiments a convection cell with very small layer thickness $d=(710\pm7)\mu$~m could be realized together with quite large lateral dimensions [($42\times 21$)$d^2$]. Thus, the dimensions of the convection cell are  such that the experiments are expected to be well described in simulations by periodic boundary conditions in the plane. Furthermore, such dimensions give a vertical diffusion time $t_{v}=d^2/\kappa=3.0$~s, which sets the time scale, is small enough for typical experiments. In the rest of this section, we will present the generic features of the patterns when all transients have died out. To compare with theory, we have performed numerical simulations  of the basic equations  (\ref{eq:nondim}) on a large horizontal domain with lateral dimension up to $L =20 \lambda_c = O(40 d)$. For the numerical method, see Appendix \ref{App:B}.

Excerpts of the experimental shadowgraph pictures to be discussed in this section  are already shown in figure  \ref{fig:3}. We will follow the same sequence of parameter combinations used 
in the preceding section, where $\gamma$ was systematically
increased. As discussed in \S\ref{sec:secstab}, we  
show in this section the vertical  temperature  average  $\langle\theta(\bm x)\rangle$ side by side with the corresponding experiments. A quantitative agreement
between theory and experiment is not to be expected, as along with the complicated optics involved in shadowgraphy \citep{Trainoff2002}, the experimental
pictures are typically  digitally remastered to enhance their contrast.

\subsection{Convection pattern for $\gamma < \gamma_{c2}$}

%%%%% figure 12
\begin{figure}
\centering{
{\includegraphics[width = 4.1cm]{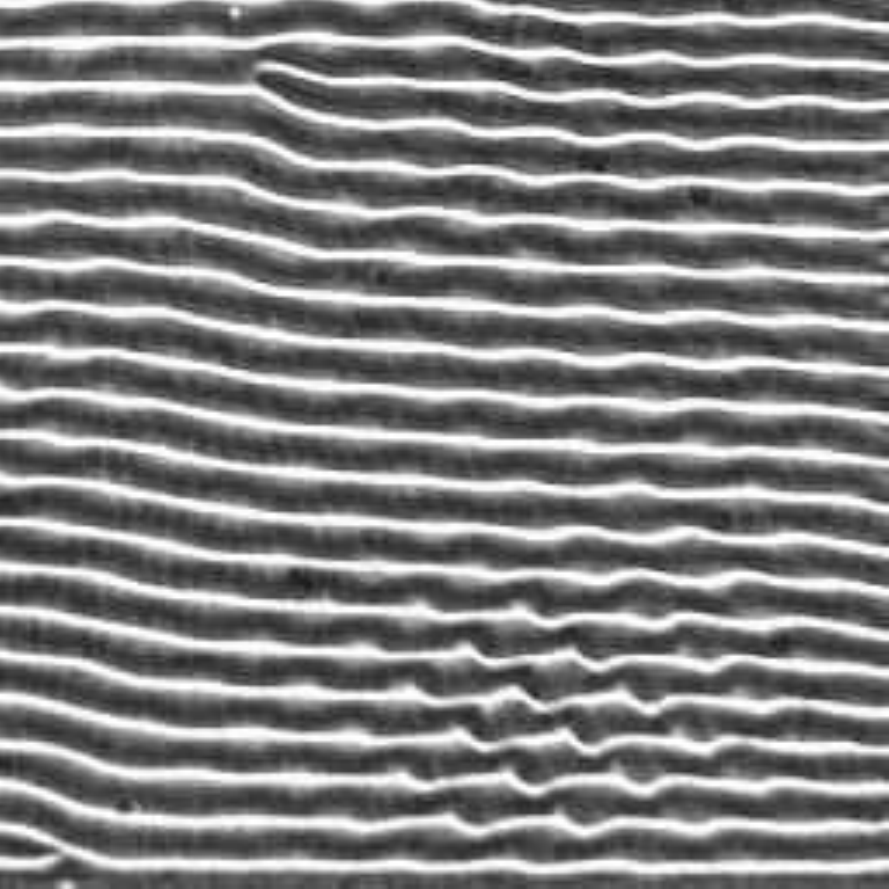}}
\hspace{1cm}
{\includegraphics[width = 4.1cm]{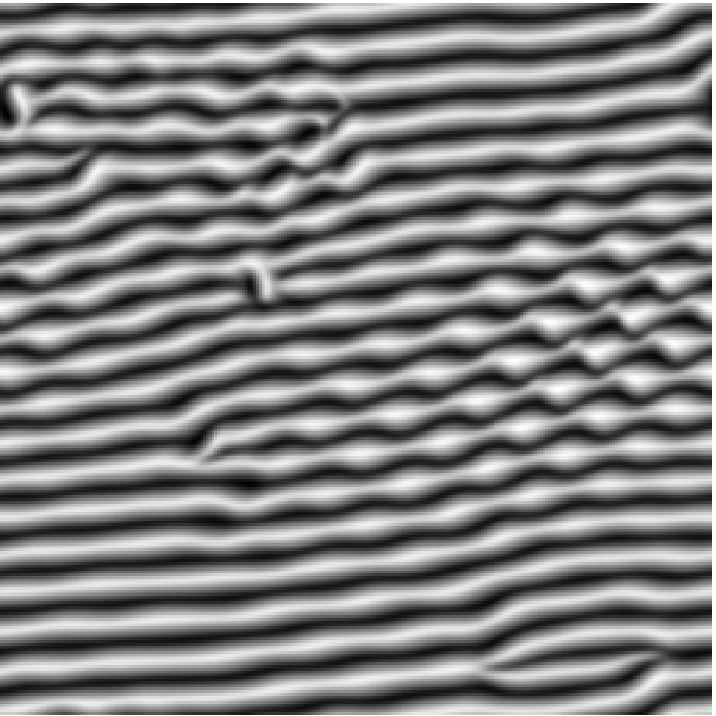}}
}
\caption{Snapshots of an longitudinal subharmonic oscillations (LSO) for $Pr=1.07$,
$\gamma=17^{\circ}$ and
$\epsilon=1.5$ from experiments (left panel) and from our numerical
simulation
(right panel).}
\label{fig:subharmonicT}
\end{figure}

We start with the subharmonic oscillatory patterns (LSO), in figure 
\ref{fig:busse} which bifurcate from longitudinal rolls. As shown in figure \ref{fig:subharmonicT}, experiments and theory match very well. We observe patches of subharmonic oscillations, which we have previously discussed in  \S\ref{sec:subhar}
 on the basis of a stability analysis of the longitudinal rolls  and numerical simulations in a minimal domain (containing only one roll pair). In large ILC
systems, typically such motifs appear only in localized patches that
compete with moderately distorted rolls. Such patches  expand and shrink
in time and their
centers move erratically over the plane. The subharmonic
oscillations within the localized patches  show an internal dynamics   with a
time scale of 1 to 3 cycles per $t_{v}$ which is of the same order as the period
$2 \pi/\omega_{inst}$  of the Hopf bifurcation in  \S\ref{sec:subhar}.

%%%%%%% figure 13
\begin{figure}
\centering{
{\includegraphics[width = 4.1cm]{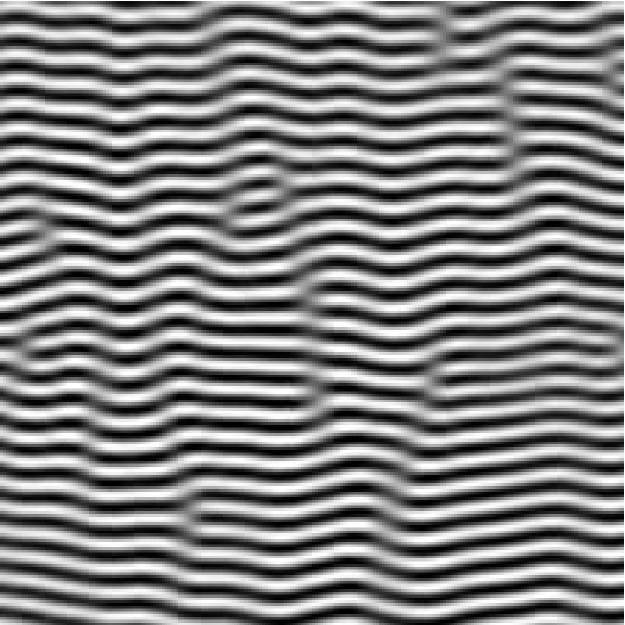}}
\hspace{1cm}
{\includegraphics[width = 4.1cm]{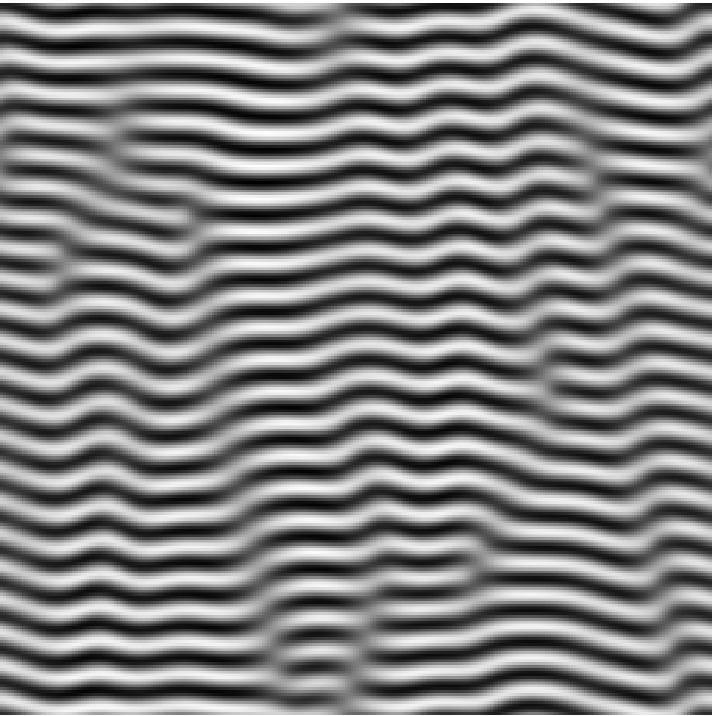}}
}
\caption{Representative  wavy roll  patterns (WR) from experiments (left panel)
in comparison 
with simulations (right panel) for  $Pr=1.07$,
$\gamma=30^{\circ}$ and $\epsilon=0.08$.}
\label{fig:ucT}
\end{figure} 

With increasing $\gamma$, the longitudinal rolls become unstable
against undulations even for very small $\epsilon$; i.e. the  instability line
in figure  \ref{fig:3} bends dramatically down. Typical experimental and
theoretical pictures of undulated (wavy) rolls (WR) shown in figure 
\ref{fig:ucT} again match very well with each other. 
However, instead of the stationary undulations as predicted in \S\ref{sec:wavy},
the patterns are  characterized by
patches of uniform undulated rolls, separated by grain boundaries
\citep{Karen2000,Karen2002}. In
addition, the rolls are scattered with point defects that move at right angles to
the rolls. The wavy patterns have been discussed in detail in
\cite{Karen2008}, where also a  weakly chaotic dynamics of the amplitudes $A,B$
in (\ref{eq:patwavy}) is analyzed in detail.

%%%%%%%% figure 14
\begin{figure}
\centering{
{\includegraphics[width = 4.1cm]{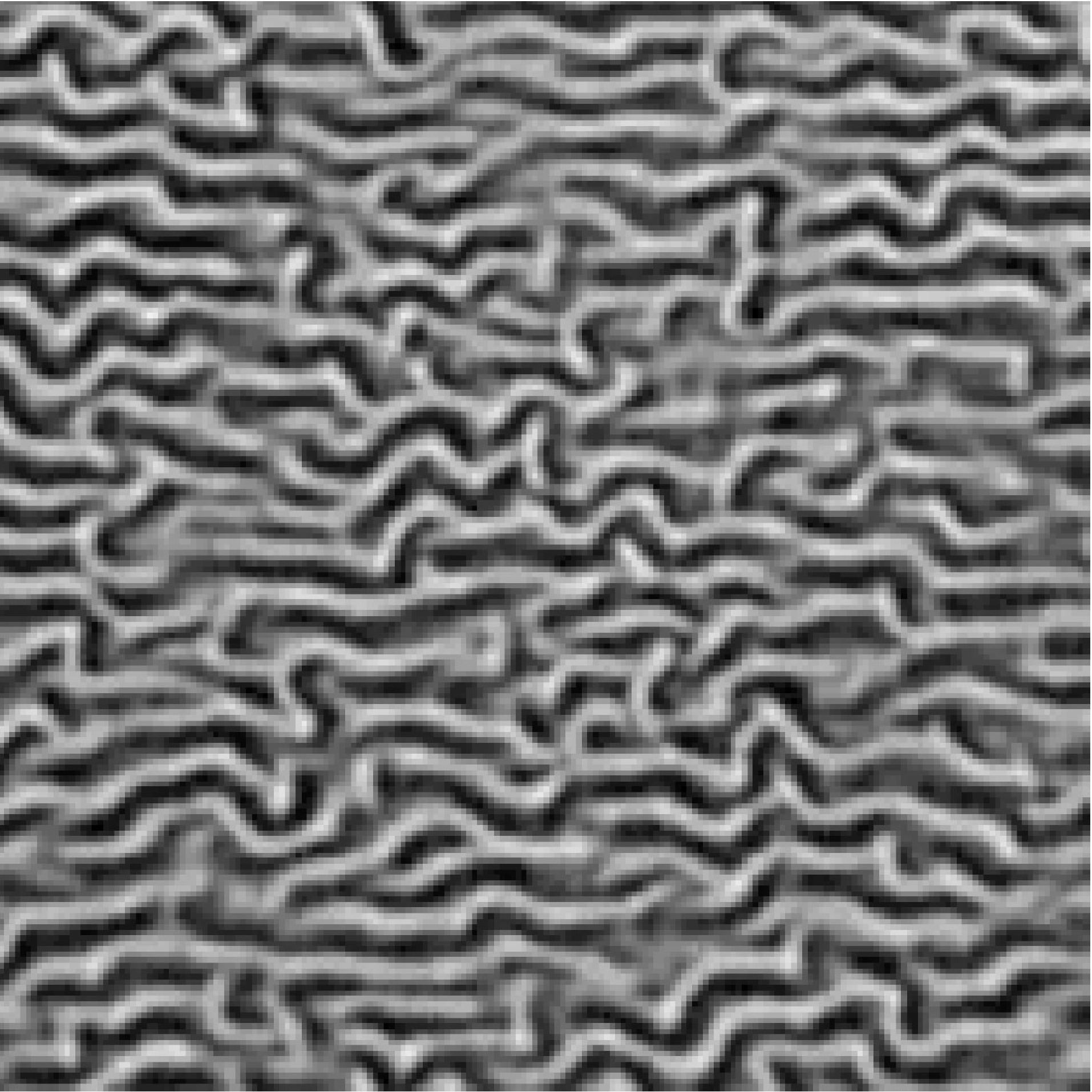}}
\hspace{1cm}
{\includegraphics[width = 4.1cm]{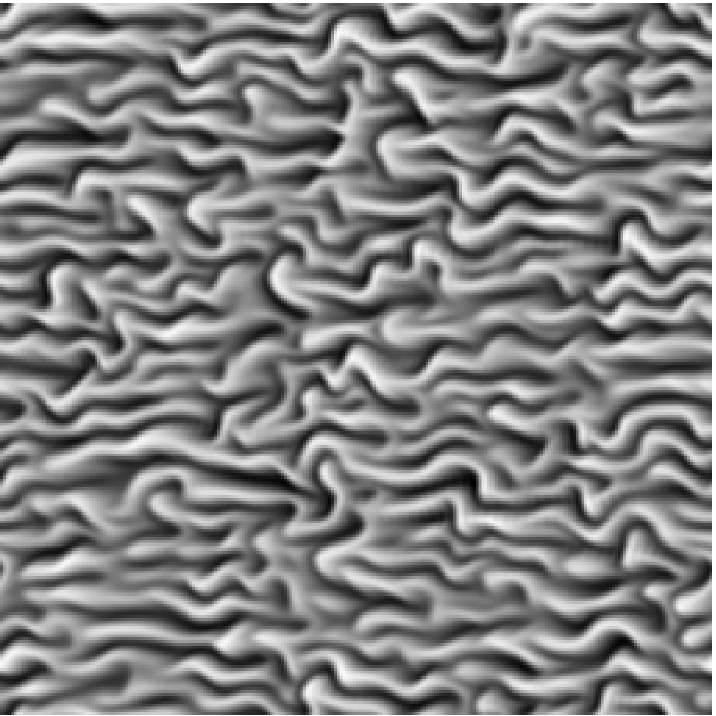}}
}
\caption{Crawling rolls (CR) from experiments
(left panel)  and from simulations (right panel) at  the same parameters as
in figure 
\ref{fig:ucT} except at a larger $\epsilon=0.4$.}
\label{fig:craw}
\end{figure}

With increasing $\epsilon$, the 
undulations become more and more disordered and the
rolls get disrupted. A transition is observed to the dynamic state of the so called crawling rolls (CR) \cite{Karen2000} as shown in figure \ref{fig:craw}. This state is also obtained in our numerical simulations, which indicates that it is not  caused by experimental imperfections. 

%%% figure 15
\begin{figure}
\centering{
{\includegraphics[width = 4.1cm]{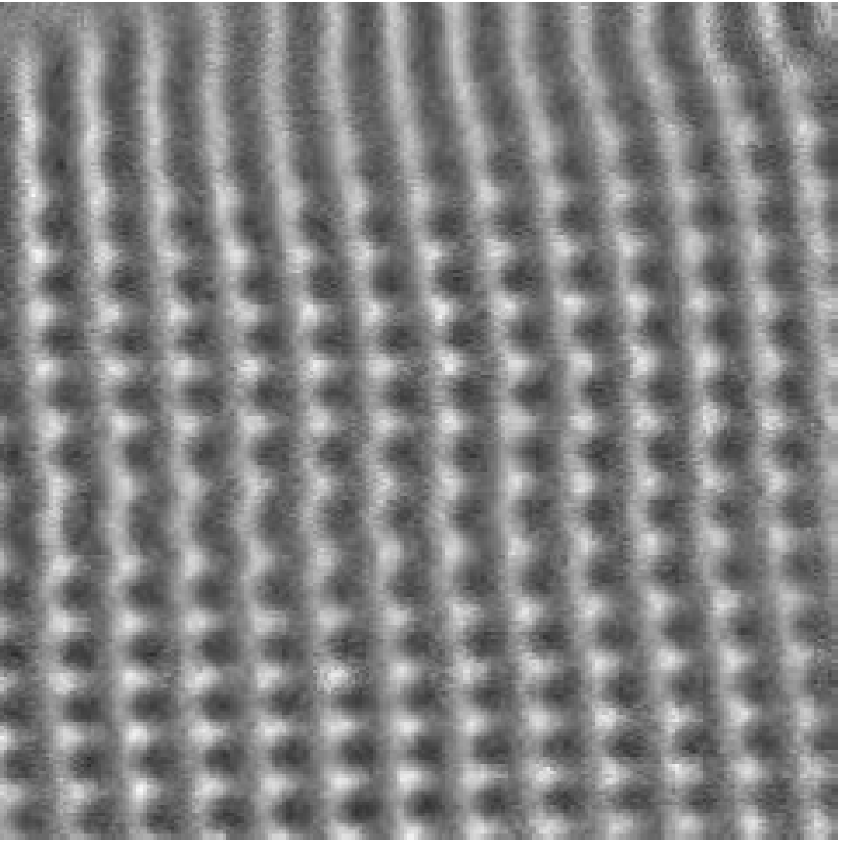}}
\hspace{1cm}
{\includegraphics[width = 4.1cm]{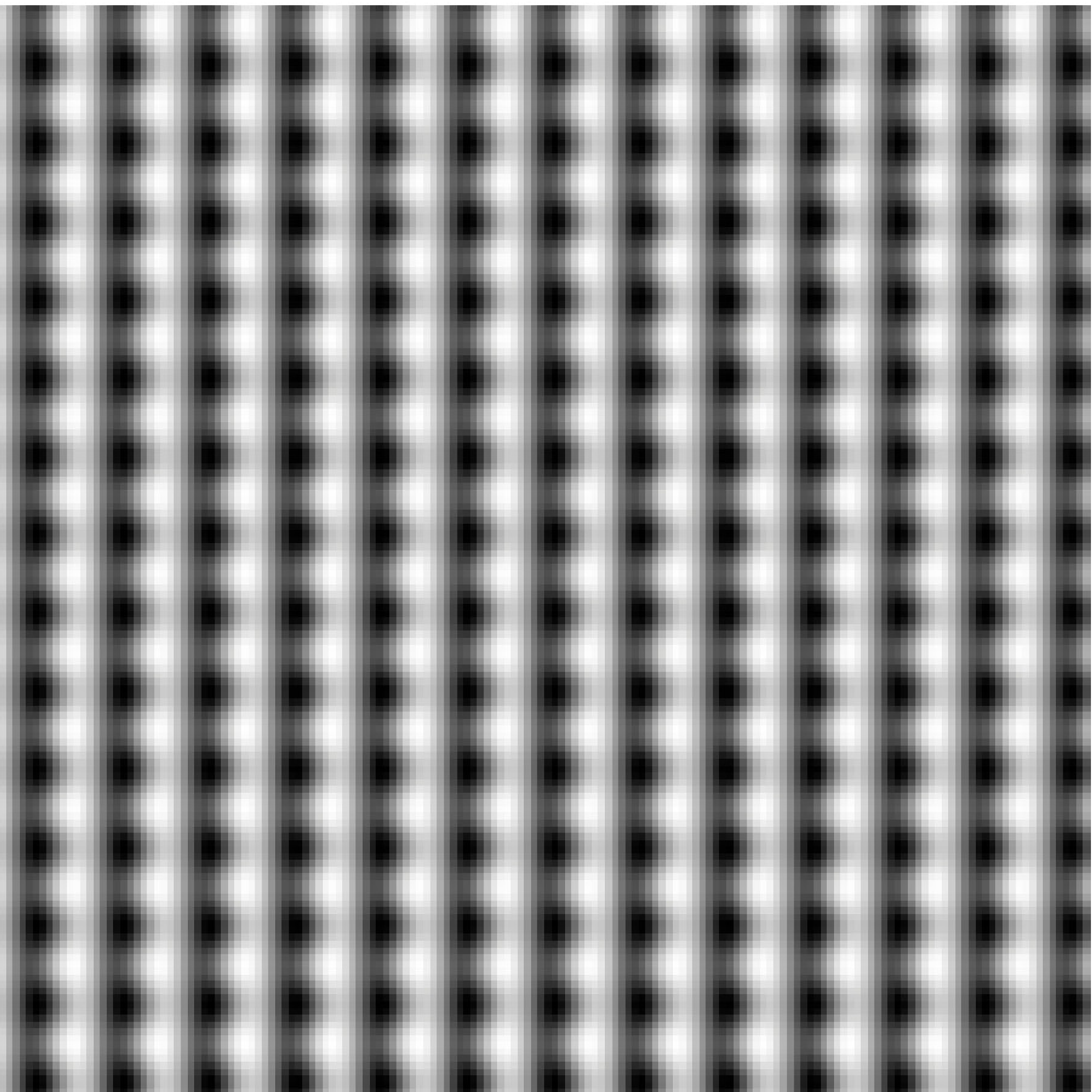}}
}
\caption{Knot pattern (KN) at
$\gamma =
80^{\circ}$ and $\epsilon=0.05$ in experiments (left panel) and in simulations
(right panel). }
\label{fig:bimod}
\end{figure}

%%%%% figure 16
\begin{figure}
\centering{
{\includegraphics[width = 4.1cm]{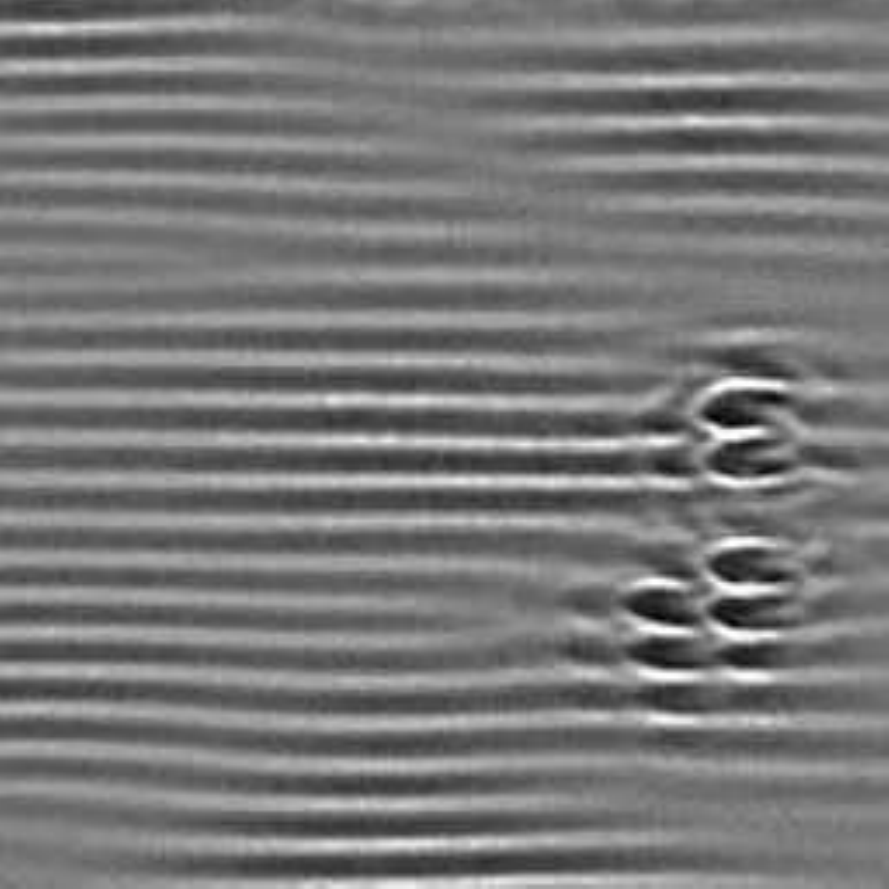}}
\hspace{1cm}
{\includegraphics[width = 4.1cm]{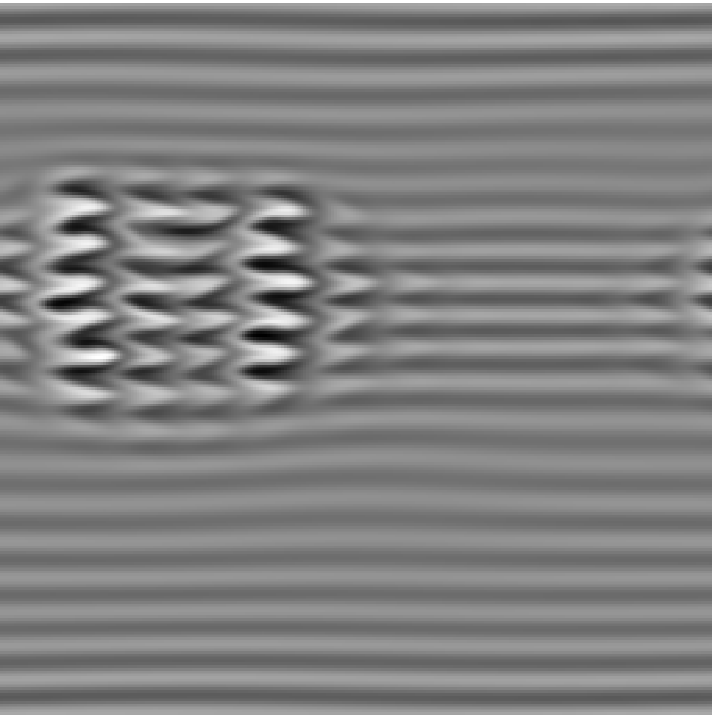}}
}
\caption{Localized transverse bursts (TB) for the parameters $Pr=1.07$,
$\gamma=77^{\circ}$ and $\epsilon=0.04$
in experiments (left panel) and in simulations (right panel).}
\label{fig:transburstsT}
\end{figure}

%----------------------------------------------------------------------------
\subsection{Convection close to codimension 2 point}
The vicinity of the codimension 2 point $\gamma_{c2}$ 
is of particular interest. According to figure \ref{fig:3} the wavy roll instability governs the secondary instability of the longitudinal rolls up to $\gamma = \gamma_{c2}$. In contrast, for $\gamma \gtrsim  \gamma_{c2}$ the primary transverse rolls are predicted to  become unstable against cross rolls leading to the knot patterns (KN). This instability mechanism is confirmed by the pictures shown in figures \ref{fig:bimod}. It is remarkable that the DNS on a large domain in the plane ($L_x = L_y = 20 \lambda_c$) starting from random initial conditions has led to perfect knot patterns. They are indistinguishable from those shown in figure \ref{fig:knot}  generated on  a small domain with $L_x, L_y \approx \lambda_c$. The teeth-like structure on the transverse rolls caused by a resonant interaction of the three roll modes (see figure \ref{fig:knot}) is born out in the experimental picture. However, the transverse rolls are slightly oblique here and undulated. 

We now discuss two types of patterns which do not allow for a direct
interpretation by secondary instabilities of the basic rolls. 
First, we show in figure  \ref{fig:transburstsT} the  transverse 
bursts (TB) for $\gamma = 77^{\circ} \lesssim \gamma_{c2}$ and at
$\epsilon=0.04$ slightly
above the secondary  wavy bifurcation  of the longitudinal rolls. Both in experiments
and simulations, we observe a background of slightly undulated rolls with
some amplitude modulations. Intermittently, localized transverse
structures (bursts) appear, which contract, vanish and reappear at other
places. The longitudinal bursts in experiments have been analyzed  in 
\cite{Karen2003}, to which we refer for more details. 

\begin{figure}
\centering{
{\includegraphics[width = 4.1cm]{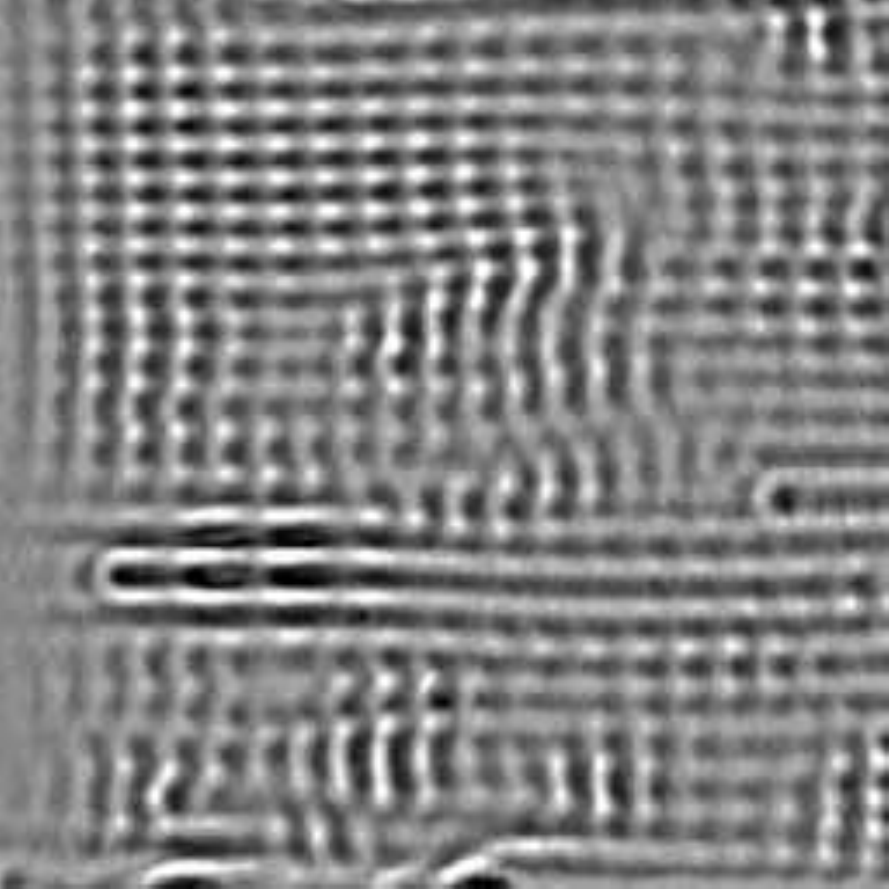}}
\hspace{1cm}
{\includegraphics[width = 4.1cm]{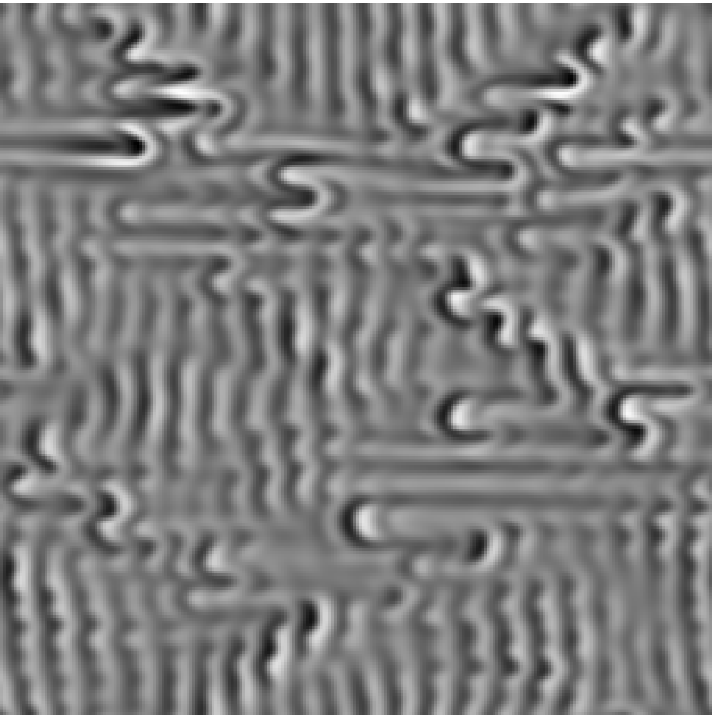}}
}
\caption{Localized longitudinal  bursts (LB) for the parameters $Pr=1.07$,
$\gamma=79^{\circ}$ and $\epsilon=0.1$
in experiments (left panel) and in simulations (right panel)}
\label{fig:longburstsT}
\end{figure}

In contrast, for $\gamma > \gamma_{c2}$  and intermediate $\epsilon$, longitudinal bursts (LB) are observed; representative examples are shown in
figure \ref{fig:longburstsT}. 
They are characterized by localized loops of longitudinal rolls superimposed on transverse rolls.
The experimental picture shows more of the bimodal knot pattern
in the background than do the
simulations. In the vicinity of  $\gamma_{c2}$,  we do not expect simulations to reproduce all details of the the experiments at the same parameters, as the system is very sensitive against small changes of $\gamma$ and $\epsilon$. The material parameters in the experiments certainly have some inaccuracies. In addition, non-Boussinesq effects presumably lead to a slow drift of the experimental pattern. Although the two burst phenomena are clearly reflected in our  simulations, additional efforts are necessary in the future to understand
their underlying mechanism.

%%%%%% figure 17

%%%%%% figure 18
\begin{figure}
\centering{
{\includegraphics[width = 4.1cm]{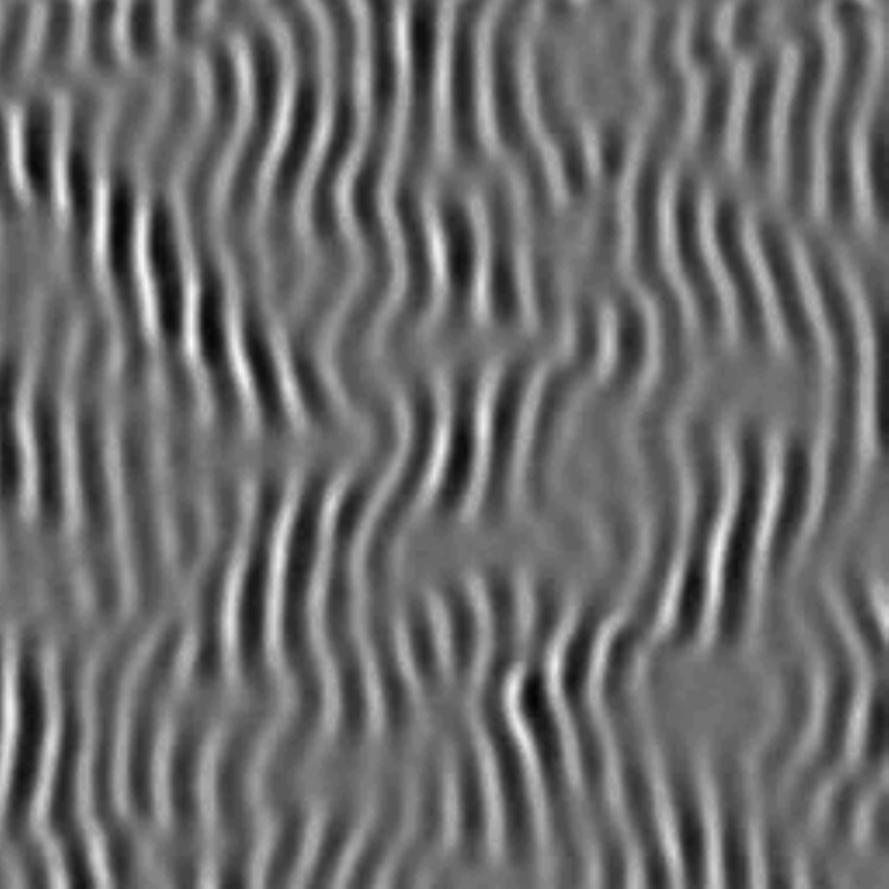}}
\hspace{1cm}
{\includegraphics[width = 4.1cm]{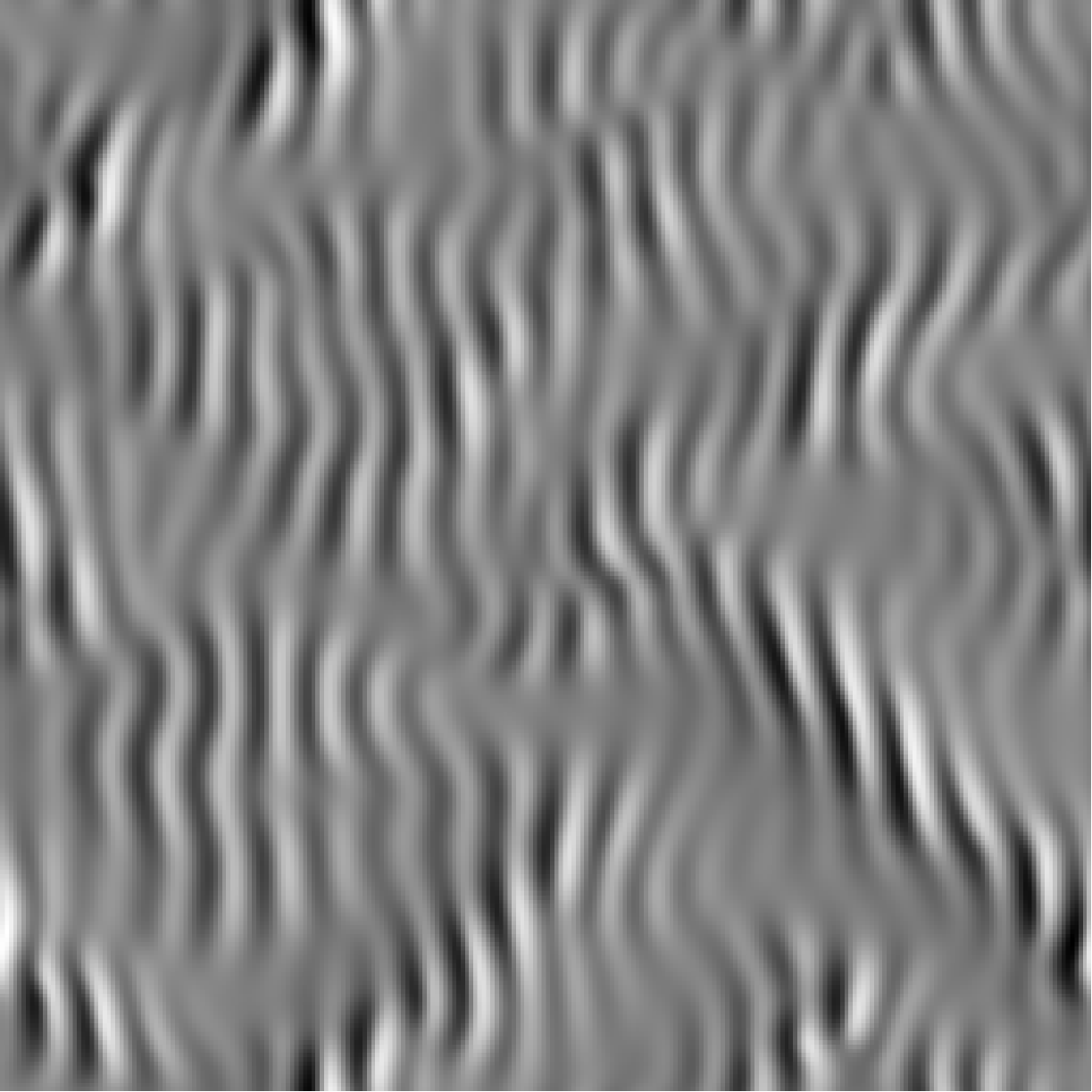}}
}
\caption{Switching diamond panes (SDP) for the
parameters $Pr=1.07$, $\gamma=100^{\circ}$ and $\epsilon=0.1$
in experiments (left panel) and in simulations
(right panel).
}
\label{fig:sdpT}
\end{figure}

%%%%%%% figure 19
\begin{figure}
\centering{
{\includegraphics[width = 12cm]{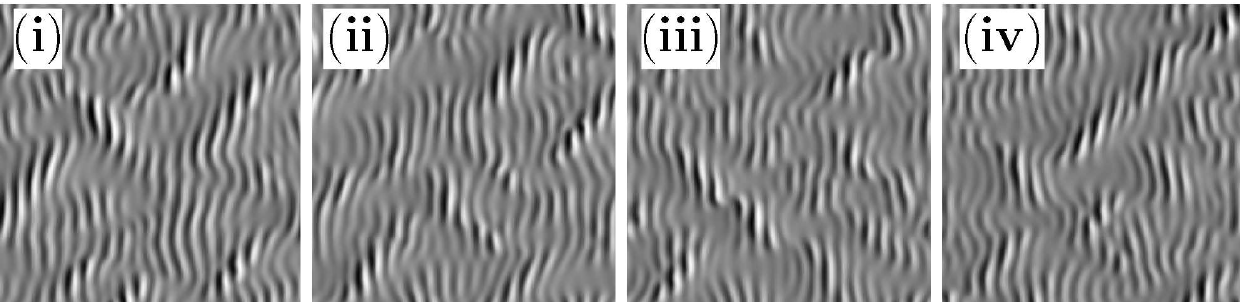}}
}
\caption{Patterns formed during a typical evolution of switching oscillatory
rolls shown in order, for the case of heating from above as in figure \ref{fig:sdpT}. Consecutive panels are separated by $5\,t_{\nu}$.
System parameters are $Pr=1.07$, $\gamma=100^{\circ}$ and $\epsilon=0.1$.}
\label{fig:sdptime}
\end{figure}

\subsection{Shear stress dominated instabilities}
\begin{figure}
\centering{
{\includegraphics[width = 4.1cm]{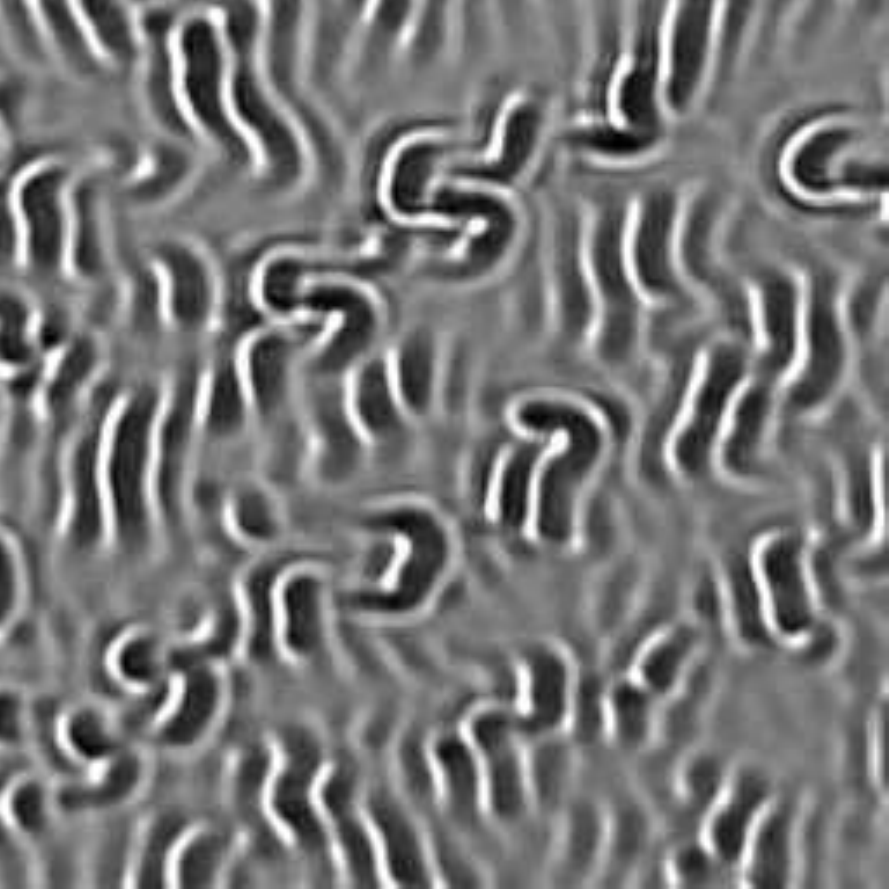}}
\hspace{1cm}
{\includegraphics[width = 4.1cm]{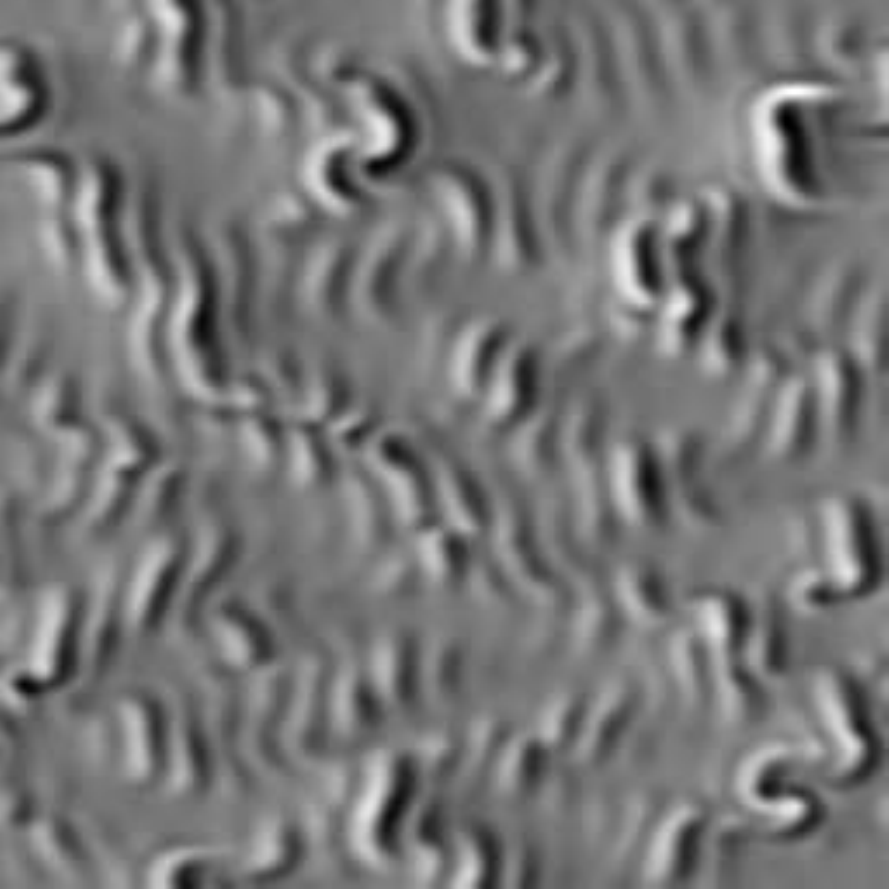}}
}
\caption{Chaotic switching rolls at 
$\gamma=100^{\circ},
 \epsilon=0.19$ [experiments (left panel),  simulations (right panel)].
}
\label{fig:sdp019}
\end{figure}

Finally, we briefly address the heating-from-above case
which is described in \S\ref{sec:rolls}, for which the inclination 
angle is $\gamma > 90^{\circ}$. As discussed before,
the destabilization of the basic state is due to the shear stress of the cubic
flow profile  $\bm U_0$ (\ref{eq:basest}).
In \S\ref{sec:sdp} we have described the destabilisation
of the primary transverse rolls to switching  diamond pane patterns shown in
\cite{Karen2002}. According to figure \ref{fig:3}, the transition line to the TO is almost
horizontal and begins at $\gamma$ slightly above $\gamma_{c2}$. 
In figure \ref{fig:sdpT}, we show a representative example for $\gamma =
100^{\circ}$, where experiment and simulations agree very well.
The time evolution observed in the corresponding  time sequence, presented in
figure \ref{fig:sdptime}, reflects the frequency $\omega_{inst} = 1.48$
given in  \S\ref{sec:sdp} and documented in figure \ref{fig:sdptime2}. Increasing $\epsilon$ further causes the patches with enhanced amplitude to get smaller and to move more erratically as shown in figure \ref{fig:sdp019}.
%%%%%% figure 20

%%%%%%%%%%%%%%%%%%%%%%%%%%%%%%%%%%%%%%%%%%%%%%%%%%%%%%%%%%%%%%%%%%%%%%%%%%%

\section{Conclusions}\label{sec:disc}

The recent experimental study  of ILC for $\Pran =1.07$   by \citet{Karen2000}
has opened a new  path to  a much better understanding of this dynamically rich 
system. For a recent overview,  see chapter 7 in \cite{Lappa2009}. In contrast to previous work, the convection
instabilities of the basic state, sketched in
figure  \ref{fig:1}, have been systematically explored as  function
of the inclination angle $\gamma$ and the Rayleigh number $R$. Furthermore, the
resulting patterns are visualized directly and the large variety of new pattern types are shown in the phase diagram in figure \ref{fig:3}. 

For the theoretical analysis in this paper, the large lateral extent of the
convection cell in the experiment has been of particular importance.  For large aspect ratio systems the influence of the lateral boundaries of the cell is certainly highly
suppressed, making it appropriate to use periodic boundary conditions. A convincing  agreement with experiments by \citet{Karen2000} has been obtained. The analysis of the primary
bifurcation of
rolls at the onset of convection and their secondary bifurcations  have revealed the complicated interaction  of buoyancy and shear driven
destabilization mechanisms.  Of particular importance is the
spatially resonant interaction between three roll systems with different
orientations in the plane (wavevector resonance, as detailed in
\S\ref{sec:secstab}).  
 
A look at the experimental pictures  and the pattern dynamics (see \cite{Karen2000}) shows that they are not completely described by perfect periodic patterns either in one or two dimensions in the fluid
layer plane.  One finds cases where a kind of clear periodicity
is expressed only in parts of the cell (figures \ref{fig:subharmonicT},
\ref{fig:ucT}).  One also observes defect lines;  in
addition the patterns change in time.
Further increasing $R$ in these cases often leads to turbulent patterns (see figures \ref{fig:craw}, \ref{fig:sdp019}). 
In addition, there are other cases where
localized patches of a different structure than the underlying, regular background patterns appear intermittently.
Two examples are the transverse bursts in figure \ref{fig:transburstsT} and the longitudinal bursts in
figure \ref{fig:longburstsT}. The ability to reproduce such weakly turbulent patterns in direct numerical simulations of the OBE validates their generic character, independent of experimental conditions.

To unravel the basic underlying mechanism is a difficult task; here we have been unable to understand these states in terms of instabilities of the underlying roll patterns. Phenomena where complex patterns that cannot be explained in terms of the instabilities of the underlying simpler patterns, have been previously described in standard RBC. A prominent example is the spiral defect chaos \citep{Bodenschatz2000}, which is often observed for medium $\Pran$ and for  Rayleigh numbers $R$ slightly larger than $R_c$, where rolls are linearly stable.

Further effort is thus 
needed  to analyze and to quantify the 
dynamics of the turbulent events in detail as 
has been done for 
the bursts in \cite{Karen2002,Karen2003} or
for the wavy patterns in \cite{Karen2008}.
Another issue is the  weakly turbulent convection states
described by Busse and coworkers in ILC, appearing even for very small systems containing only one roll pair
\citep{Busse1992,Busse2000}.  Their relation to the weakly turbulent events,
which here cover considerably larger domains in the plane,  requires further
investigation. 
 
In this paper, we have restricted ourselves to the
special case of $\Pran =1.07$. As part of future work, it is planned to apply
our methods in particular to fluids with large $\Pran > 12.47$,
 where the primary roll bifurcation is oscillatory.

\begin{acknowledgements}
The authors are highly indebted to Prof. F. Busse for his very useful comments and
fruitful discussions on the subject of this paper.
\end{acknowledgements}

\appendix

\section{Governing equations and stability of rolls}\label{App:a}
%%%%%%%%%%%%%%%%%%%%%%%%{\color{red} Equations and Oberbeck-Boussinesq assumption}\\

In section \S\ref{sec:gde}, the poloidal-toroidal decomposition of the
solenoidal velocity field $\bm v $ (\ref{eq:reduc}) is 
written as follows:
\bee
\label{eq:velpot}
\bm v(x,y,z,t) =  \bm \nabla \times (\bm\nabla \times f\hat{\bm z}) +
{\bm\nabla} \times \Phi {\hat{\bm z}} + \bm U(z,t) \equiv {\bm \chi}f+ {\bm
\eta} \Phi + \bm U(z,t).
\ee 
The explicit equations for $\theta, f, \Phi$ are obtained by
inserting  (\ref{eq:reduc})  into 
(\ref{eq:nondim}) followed by requiring the divergence of velocity to vanish in (\ref{eqn:nondimU}).
The evolution equation for the secondary meanflow flow
$\bm{U}(z,t)$ is obtained by averaging the velocity equation 
(\ref{eqn:nondimU}) over the
$x-y$ plane, leading to:
\bee
\label{eq:meanU}
\frac{1}{\Pran}\frac{\partial \bm U(z,t)}{ \partial t}  =
-\frac{1}{\Pran}\frac{\partial \overline{(v_z \bm v)}}{\partial z}
  + \frac{\partial^2 \bm
U}{\partial
z^2}  + \sin\gamma\, \overline{\theta} -(\partial_x, \partial_y,0) (P_x  x  +P_y
y),  
\ee
where the  overbar indicates  a  horizontal average.

Except for minor changes, the resulting equations for $\theta, f, \Phi$  can already be
found in \citet{Busse1992, Karen2008}. The differences arise firstly from the definition of the Rayleigh number $R= \Delta T \cos \gamma /T_s$, whose explicit dependence on $\cos \gamma$, is not convenient for the description of vertical convection cells when $\gamma =90^0$. By the transformation $\theta \rightarrow \theta/ \cos \gamma$ and interchanging $x$ and $y$ we arrive at our formulation. Secondly, our equations for the meanflow $\bm U$ (\ref{eq:meanU}) contain the additional pressure terms $P_x(t), P_y(t)$.   They have  been  proposed in a different context in \citet{Busse2000}, to guarantee mass conservation, $\int dz \bm U(z) =0$. Finite $P_x(t), P_y(t)$  appear only in the DNS of complex patterns in \S\ref{sec:comp}.

For the following discussions, a compact symbolic representation of the
equations for the fields $\theta, f, \Phi$  is useful:
\begin{eqnarray}
\label{eq:sym1}
\widehat {\cal C} \; \frac{\partial}{\partial t}
\Vhat{V} (\bm x, z,t) = \widehat {\cal L}  \widehat{\bm V}(\bm x, z,t) +
\widehat{\bm
N}[\widehat{\bm V} + \bm U, \widehat{\bm V}]
\end{eqnarray}
with $\bm x = (x,y)$ and the symbolic vector $\Vhat{V} = [\theta , f, \Phi]^T$. 
The symbol $\widehat{\bm N}$  stands for the nonlinear terms which consist of
quadratic forms in $\theta, f,\Phi$ and $\bm U$.  

As an example, we show the explicit expressions for the linear terms of $\theta$ and $f$. This allows us to immediately identify the corresponding  components
of the linear operators $\widehat {\cal C},\widehat{\cal L}$:
\begin{subequations}
\label{eq:incfgh}
\beal
\frac{\partial}{\partial t} \theta  &= -R \Delta_2 f + \nabla^2 \theta
- R \sin \gamma (U^x_0(z) \partial_x)
\theta, \label{eq:linthzet1n}\\
\frac{1}{\Pran} \partial_t \nabla^2 \triangle_2 f &=\nabla^4 \triangle_2 f
-\cos \gamma\, \triangle_2 \theta  + \sin \gamma\, \partial_x \partial_z 
 \theta   -\frac{1}{\Pran}  \sin \gamma  R \, F[U^x_0] f\,
\label{eq:incfn}
\end{align}
\end{subequations}
with $\Delta_2 = (\partial_{xx} + \partial_{yy})$. The term
$F[U^x_0]  \equiv  [U^x_0(z) \nabla^2  -
\partial^2_{zz}
U^x_0(z)] \partial_x \Delta_2 $
originates from the contribution of the basic mean flow $\bm U_0$ in
 (\ref{eq:basest})  to the velocity $\bm u$   in
(\ref{eqn:nondimU}). Note that $\theta, f$ are not coupled to $\Phi$ in
(\ref{eq:incfgh}). 

In general, equations (\ref{eq:sym1}) are solved with the boundary conditions $\theta(z = \pm 1/2) =0$ and
$f=\partial_z f = \Phi= \bm U =0$ at $z =  \pm 1/2$ which derive from the 
 no-slip boundary conditions $\bm v(z = \pm 1/2) =0$. These conditions are automatically satisfied by the use of Galerkin expansions with respect to $z$. As in \citet{Busse1992}  we use for $\theta$ 
the ansatz:
\begin{equation}
\label{eq:galerk}
\theta(\bm x, z,t) = \sum_{m =1}^{M} S_m(z) {\vartheta}_m(\bm x,t);\qquad
S_m(z) = \sin (m \pi (z+1/2)),
\end{equation}
since $S_m(z = \pm 1/2) =0$.  For  $\Phi$ and
the secondary mean flow $\bm U(z,t)$ in (\ref{eq:meanU}) also
sine functions are used, while $f$ is expanded in terms of the Chandrasekhar
functions $C_m(z)$ \citep{chandra1961} with $C_m(\pm 1/2) = \partial_z C_m(z =
\pm 1/2) =0$.

\subsection{Linear Stability Analysis of the basic state}
\label{App:alin}
The primary convection instability of the basic state corresponds to  
exponentially growing solutions in time of  (\ref{eq:sym1}) in the linear
regime $(\widehat{\bm N} =0)$.
We use the ansatz  $\widehat{\bm V} (\bm x, z, t) = 
e^{\lambda t}  e^{ \im \bm q \cdot \bm x} \tilde{\bm V} (\bm q, z, R)$ in
(\ref{eq:sym1}) to arrive at the following linear eigenvalue problem for
$\sigma$:
\begin{equation}
\label{eq:eig1}
\sigma {\cal {C}}(\bm q,\partial_z) \tilde{\bm V}(\bm q, z;R) = {\cal L}
\tilde{\bm V}(\bm q,z;R) \equiv  [{\cal
A}(\bm
q,\partial_z) + R {\cal B} (\bm q, \partial_z)] \tilde{\bm V},
\end{equation}  
where the operators ${\cal{C,L}}(\bm q,\partial_z)$  etc. in Fourier space
derive
from the corresponding ones in position space (see  (\ref{eq:sym1})) carrying a hat symbol via the
transformation
$\partial_{\bm x}  \rightarrow \im \bm q$. In this paper  we make use of Galerkin expansions ((see (\ref{eq:galerk})) to handle the $z$ dependence.   Thus, for instance (\ref{eq:incfgh}) is transformed into an algebraic linear eigenvalue problem of dimension $2 M$ in the Fourier-Galerkin space.

If $\sigma_{max}(R,\Pran,\gamma, \bm q)$ is the eigenvalue with the largest real
part in  (\ref{eq:eig1}), then rolls become unstable when  $\sigma_{max}(R,\Pran,\gamma, \bm q)$ crosses zero. The standard procedure to determine the neutral surface $R = R_0(\Pran,\gamma;\bm{q})$ through the condition $\sigma_{max}(R =R_0,\Pran,\gamma, \bm q)=0$. The minimum of $R_0(\bm{q},\gamma)$ with respect to $\bm{q}$ gives the critical 
wavevector
$\bm{q}_c$  and  the critical Rayleigh number $R_c = R_0(\bm q_c)$ as
function of 
$\Pran, \gamma$. If the frequency $\omega_c \equiv
\Imag[\sigma_{max}(R_c, \bm q_c)] =0$, the bifurcation of the basic state is
${\it stationary}$ otherwise it is ${\it oscillatory}$. This method works in the ranges $\gamma \lesssim 12^0$ and $\gamma \gtrsim 17^{0}$ in figure \ref{fig:3}, where the rolls are stable for $R <  (1 + \epsilon_{inst}) R_c(\gamma)$. Between $12^{\circ} < \gamma <  17^{0}$, where $R_0 (\bm q, \gamma)$ is not unique, we analyze $\sigma_{max}(R,\Pran,\gamma, \bm q) =0$ using scans of $\gamma, \bm q$ at fixed $R$ to determine the upward-bent threshold of the wave roll instability, which limits the stability of the longitudinal rolls.

It turns out, that eigenvalues $\sigma$ with  $\Real [\sigma_{max}]
\ge 0$ are obtained by reducing the eigenvalue problem  (\ref{eq:eig1})
to its $\theta, f$ part (\ref{eq:incfgh}). 
Thus, the use of our  Galerkin-expansions (\ref{eq:galerk}) reduces Eq. (\ref{eq:eig1}) to an algebraic linear eigenvalue problem with $2
M \times 2 M$ matrices, which is analyzed using standard linear algebra codes (LAPACK).   

\subsection{Secondary  of instabilities of roll solutions}\label{App:asecond}

For intermediate $\Pran$, considered in this paper,  the primary bifurcation to rolls
with wavevector $\bm q_c$  at $R = R_c$ is stationary. To construct the
evolving finite-amplitude solution $\Vhat{V}  = \Vhat{V}_r$  for $R > R_c$ from
(\ref{eq:sym1})  we use 
the Fourier ansatz:
\begin{equation}
\label{eq:four2d}
\Vhat{V}_r (\bm x,z)=  \sum_{k = -N/2}^{n = N/2}  e^{i k \bm q_c
\cdot \bm x}
\bm V_r( k \bm
 q_c, z). 
\end{equation} 
With respect to $z$, we introduce an additional Galerkin expansion (see (\ref{eq:galerk})) of the $(N +1)$ Fourier coefficients
$\bm V_r(k \bm q_c,z)$. Furthermore, the Galerkin expansion of the mean-flow $\bm U$ using sine functions leads to $2 M$ additional equations.      
Thus, we arrive at a system of $3 M (N+1) +2 M $  coupled nonlinear
algebraic equations for all Galerkin expansion coefficients. This system is solved by
Newton-Raphson methods. 
 
The iteration process is started from the weakly-nonlinear roll solution 
 of  (\ref{eq:sym1})  characterized  in Fourier space by the ansatz  
$\bm {V}_{wnl}(\bm q_c,z)= A(\bm q_c, R)  \bar{\bm V}_{max} ( \bm{q_c},z; R)$;
 $\bar{\bm V}_{max}$ is given by the solution of  (\ref{eq:eig1}) for $\sigma = \sigma_{max}$ at $\bm q = \bm q_c$. For $\epsilon \gtrsim 0$, a systematic expansion with respect to the small parameter $\epsilon$ determines the amplitude $A$ of $\bm V_{wnl}$ via the solution of the amplitude
equation:
\bee
\label{eq:amp} 
\sigma_{max}(\bm q_c, R, \Pran, \gamma) A  -  c A |A|^2 =0, \quad
\textrm{with} \quad Re[\sigma_{max}] \propto \epsilon\,.
\ee  
For stationary primary ILC bifurcations at intermediate $\Pran$,  both $\sigma_{max}$ and the cubic coefficient $c$ are always real and positive and the bifurcation is thus forward (supercritical) at  onset, i.e.  $|A|^2  \propto \epsilon/c$ increases continuously beyond the threshold $\epsilon =0$.

To examine  the linear stability of the roll solutions $\hat{V}_r$
(\ref{eq:four2d}), we linearize (\ref{eq:sym1}) with respect to an
infinitesimal perturbation
$\delta \Vhat{V}_r(\bm x,z,t)$ of $\Vhat{V}_r$.  
Then we arrive at a set
of coupled linear equations for the components of $\delta \Vhat{V}_r$,
which are solved by the 
standard Floquet ansatz:
\begin{equation} 
\label{eq:floquet}
\delta \widehat{\bm V}_r(\bm x, z,t) 
= e^{\Lambda \, t} e^{i \bm s' \bcdot
\bm x }
\sum_{k = -N/2}^{k = N/2}  \, e^{i  k \bm q_c \bcdot \bm x} \delta 
\bm V_r(k \bm q_c,z), 
\end{equation}   
Thus, we arrive at a linear eigenvalue problem for the eigenvalues $\Lambda(\bm
s', \bm q_c, R)$. This is mapped, as before, to a linear algebraic problem using an Galerkin expansion of  
$\delta \bm V_r(k \bm q_c,z)$. The eigenvalue $\Lambda_0$ with the largest real
part defines the growth rate $\lambda_0(\bm s') \equiv \Real[\Lambda_0(\bm s')]$
of the perturbation $\delta \widehat{\bm V}_r$.

Given that  $\lambda_0(\bm s')$  assumes its  maximum  
$\lambda_{max}(R,
\Pran,\gamma)$ at   $ \bm s'=\bm s'_{max}$, we then determine 
the smallest  Rayleigh number $R = R_{inst}(\gamma, \Pran)$   at
which $\lambda_{max}$ crosses zero for every $R$.
 In other words at $R = R_{inst}$  
the {\it secondary} instability of the rolls with wavevector $\bm q_c$
occurs for given  parameters  $\Pran,\gamma$. When $\omega_{inst} = Im[\Lambda_0(\bm
s', R_{inst})] \ne 0$ the secondary bifurcation is called oscillatory, 
otherwise stationary.  
For an appropriate interpretation of the instability, we
have to determine for a
given  $\bm s'_{max}$ the
index  $k_{max}$  corresponding to the largest modulus $|\delta \bm V_r(k_{max}
\bm q_c,z))|$ of the expansion coefficients in (\ref{eq:floquet}). This
yields then  the  wavevector(s) $\bm q_{inst}$ of the dominant 
destabilizing mode(s)  as $\bm q_{inst} = k_{max} \bm q + \bm s'_{max}$. It
turns out that 
$k_{max}$ is always governed by the temperature component of $\delta \bm V_r$
and that $|k_{max}| \le 1$ holds.

The stability of roll patterns along the method described above has been
intensively employed for the investigation of the standard
isotropic RBC ($\gamma =0$) by Busse
and coworkers also for $\bm q \ne \bm q_c$. They have constructed the
Busse balloon \citep{Busse1979,Busball1996}, which is the stability diagram of rolls
with wavevector $\bm
q$ in the $R, \Pran-$ parameter space.
In our anisotropic ILC system the calculation of the full Busse
balloon is possible if we replace $\bm q_c$ by arbitrary $\bm q$ everywhere in
equations (\ref{eq:four2d} -  \ref{eq:floquet}).

Here we consider only the special case $\bm q
= \bm q_c$, where  
only finite $\bm s'$ perturbations
have been  found to be relevant. This has been discussed
in detail in \S\ref{sec:secstab}. In particular in several cases the same
maximal value of $|
\delta \bm V(k \bm q,z))|$  is assumed at two different integers $k_2,
k_3$ 
and thus two different dominant destabilizing modes $\bm q_{inst} = \bm
q_2, \bm q_3$ exist.
The resulting patterns are in addition  often characterized by {\it wavevector
resonances} of the form $\bm q_1 + \bm q_2 + \bm q_3 =0$
between the wavevector $\bm q_1 =\bm q_c $ of the basic roll pattern and the
$\bm q_2,\bm
q_3$,
which characterize the 3D patterns for $R>R_{inst}(\bm q_c, \gamma)$ in
Fourier space. The stability analysis yields also  the
relative phases of the  three Fourier amplitudes. Because of translational
invariance of the system in the $x-$ and the $y-$ directions two phases
can be chosen to be zero without loss of generality; the third
one is then determined
by the ratio of the components of $\delta \bm V_r(k_2 \bm q_c,z)$ and
$\delta \bm V_r(k_3\bm
q_c,z)$ with the largest moduli.

\subsection{Accuracy of the stability limits}
\label{App:astabaccur}
The accuracy of all thresholds and results presented  in this paper depend on the choice of 
the truncation parameter $M$ of the Galerkin expansion with respect to $z$ (see  (\ref{eq:galerk})) and the truncation parameter $N$ in Fourier space (see (\ref{eq:four2d}, \ref{eq:floquet})). In this paper we have always chosen $M=8$ and $N=5$. By systematically increasing  these parameters (see below) we have tested that this choice is sufficient to guarantee that the relative errors of all results given in this paper are below $0.1\% $. Even reducing the values to $M =6$ and $N =3$ does not practically change the curves shown in this paper.   

In the table \ref{tab:t1} we give representative examples for linear threshold values $R_c$ and $q_c$ and their dependence upon increasing values for the Galerkin truncation parameter $M$. In addition, the determination of the codimension 2 point varies as $\gamma_{c2} = 77.7857^\circ$, $77.7462^\circ$, $77.7544^\circ$ and $77.7560^\circ$ when varying $M$ as $M=6,8,10,12$ respectively. 

\vspace{0.2cm}

\begin{table}
\begin{minipage}{0.2\textwidth}
%\begin{table}
 \begin{center}
\def~{\hphantom{0}}
 \begin{tabular}{cccc}
  $\gamma$ &  $M$ & $q_c$ & $ R_c $ \\
  \hline
$0^{\circ}$ &  6 & 3.1159 & 1707.985\\
          &  8 & 3.1162 & 1707.824 \\
          & 10 & 3.1163 &  1707.784 \\
          & 12 & 3.1163 &  1707.771\\
\end{tabular}
\end{center}
%\end{table}
\end{minipage}
\hspace{1.5cm}
\begin{minipage}{0.2\textwidth}
%\begin{table}
 \begin{center}
\def~{\hphantom{0}}
 \begin{tabular}{cccc}
  $\gamma$ &  $M$ & $q_c$ & $ R_c $ \\
  \hline
$90^{\circ}$ &  6 & 2.7920 & 8504.200 \\
           &  8 & 2.8076 & 8470.286 \\
           & 10 & 2.8059 & 8476.495 \\
           & 12 & 2.8056 & 8477.690 \\

\end{tabular}
\end{center}
\end{minipage}
\hspace{1.5cm}
\begin{minipage}{0.2\textwidth}
%\begin{table}
 \begin{center}
\def~{\hphantom{0}}
 \begin{tabular}{cccc}
  $\gamma$ &  $M$ & $q_c$ & $ R_c $ \\
  \hline
$100^{\circ}$ &  6 & 2.7723& 9150.774\\
          &  8 & 2.7985 & 9108.171 \\
          & 10 & 2.7877 &  9115.841 \\
          & 12 & 2.7871 & 9117.068 \\

\end{tabular}
\end{center}
%\end{table}
\end{minipage}
\caption{Dependence of threshold values $q_c$ and $R_c$  for $\Pran=1.07$ of
longitudinal rolls at
$\gamma=0^{\circ}$ (horizontal cell), transverse rolls at $\gamma=90^{\circ}$
(vertical cell) and at  $\gamma=100^{\circ}$ (heating from above) upon different
Galerkin truncation parameters 
$M$.
}\label{tab:t1}
\end{table}
Analogous convergence checks have been performed  for the secondary instabilities of the convection rolls. In general the data  are more sensitive  against  changes of $M$ than of $N$.  We did all calculations with $M= 8$, which guarantees the same accuracy of the data as in the linear regime above.

To guarantee an relative accuracy of better than $0.1\%$ $N =3$ is sufficient at small $\epsilon$; for $\epsilon = O(1)$ one needs $N\geq 4$. This conclusion is supported by the following representative data. The secondary instability towards wavy rolls in figure \ref{fig:3} at $\epsilon_{inst}=  0.8$ occurs at $\gamma =16.50^\circ$  for $N=5$ and at $\gamma=16.5288^\circ$,  for  $N =4$. For $N =8$ The  LSO instability at  $\gamma =17^\circ$ is characterized by $\epsilon_{inst}= 1.04423$ with the circular oscillation frequency $\omega_{inst} = 10. 20808$ and Floquet vector as $s_x = 1.27929$ and $s_y =q_{c0}/2$. With the smaller $N =4$ we find only small changes with $\epsilon_{inst}= 1.0436, \omega_{inst}=10.2038$ and $s_x = 1.278990, s_y= q_{c0}/2$. Finally we mention the knot instability of transverse rolls at $\gamma = 83^\circ$. For  $N =5$ we find $\epsilon_{inst}=0.07134$  with $s_y = 3.1179$, which remain unchanged for $N=4$.

\section{Direct simulations of the OBE in ILC}\label{App:B}
Direct simulations of the OBE in (\ref{eq:nondim}) are in general confined to a
rectangle in the $x-y$ plane with the lateral extensions $L_x,L_y$ using  
periodic boundary condition $\Vhat{V}(x, y,z) = \Vhat{V} (x+L_x, y+L_y,z)$.
Thus, we transform to Fourier space by introducing  a discrete  2D Fourier
transformation of $\Vhat{V}$ on a $N_f
\times N_f$ grid with mesh sizes $\Delta q_x, \Delta q_y$  in the $x$- and
$y$-directions:
\begin{equation}
\label{eq:four3d}
\Vhat{V}(\bm x,z,t)=  \sum_{\bm q}  e^{i \bm q \cdot \bm x}
\bm V(\bm
q,z,t)\,\, \textrm{where}\,\, \bm q = \{(k \Delta q_x, l \Delta q_y)\} \;
\textrm{with} \,
-N_f/2 \le (k,l) \le \,N_f/2\,,
\end{equation}
and $/\Delta q_x = 2 \pi/L_x, \Delta q_y = 2 \pi/L_y$.
Reality of $\Vhat{V}(\bm x,z,t)$ implies the
condition $\bm V(\bm
q) =  \bm V(-\bm q)^{*}$. With respect to $z$, we use Galerkin expansions with
the truncation parameter $M$ as before.
The quadratic nonlinearities $\widehat{N}$ in (\ref{eq:sym1}) are treated by 
standard pseudospectral methods (see e.g. \cite{Boyd2001}). 
Substituting the Fourier ansatz (\ref{eq:four3d})  into (\ref{eq:sym1}) and
projecting on the respective Galerkin modes one arrives at a system of $3 M
\times N_f^2$ coupled ordinary differential equations for the evolution of all the combined
Fourier-Galerkin expansion coefficients. In addition, (\ref{eq:meanU}) is  mapped into a system of 2 M equations for the Galerkin coefficients of the secondary meanflow $\bm U$. Semi-implicit time stepping methods, as sketched  in the following subsection, are used to compute the time evolution  of all our fields.  

For all DNS shown in section \S\ref{sec:comp}, we have used $L_x = L_y =
n_L \lambda_c$ with  the critical wavelength  $\lambda_c =  2 \pi/q_c$ and the truncation parameters $N_f = 256,  M=8$ and  $n_L =20$.  Note that the number
of roll pairs (black and white stripes) of the underlying 2D roll patterns directly reflects $n_L$.

As also evident from the previous section our Fourier coefficients decay quickly with increasing $|\bm q|$. The truncation parameter  $N_f = 256$  in the simulations corresponding to $N = (124/20) > 5$, is used for the stability analysis of rolls in Appendix \ref{App:asecond}. Thus it is not surprising that the results of the Floquet analysis are reproduced in the DNS. Increasing $N_f$ and/or decreasing $n_L$ corresponds to keeping Fourier modes with larger $|\bm q|$. We have checked that all the typical scenarios discussed in \S\ref{sec:comp} are recovered.

The goal of most simulations in \S\ref{sec:secstab} was to validate the secondary instabilities of rolls originally obtained on the basis of \ref{App:asecond}, which lead to strictly-periodic $3D$ patterns. Thus, we have performed the DNS on minimal domains in the plane, where one side length was given as $\lambda_c$ while the other was determined by the wavevectors of the dominant destabilizing modes $\bm q_2, \bm q_3$ introduced in Appendix \ref{App:asecond}.  The data have then been mapped to a larger domain in the $x-y$ plane by periodically extending the minimal domains for visualization.

%%%%%%%%%%%%%%%%%%%%%%%%%%%%%%%%%%%%%%%%%%%%%%%%%

\subsection{Exponential Time Differencing method} \label{App:btime}

Our starting point is (\ref{eq:sym1}), where  the components of $\bm V(\bm q,z,t)$ in  (\ref{eq:four3d}) are expanded into the appropriate Galerkin modes like in (\ref{eq:galerk}).  In the  resulting Fourier-Galerkin (\ref{eq:sym1}) can be written as:
\bee 
\label{eq:gl1n}
\frac{d}{dt} \bm V(t) = {\mathsf A}  \bm V(t) - \bm {\tilde N} 
\quad  \textrm{with} \quad  \bm {\tilde N} = {\cal C}^{-1}\bm N, \, {\mathsf A}
= {\cal C}^{-1}{\cal L},
\ee 
since the matrix $\cal C$ is not singular. Equation (\ref{eq:gl1n}) allows for the formal solution:
\bee 
\label{eq:solvV}
\bm V(t+dt)  = e^{{\mathsf A} dt} \bm V(t) - e^{{\mathsf A} (t + dt)} \int_t^{t +dt} e^{{-\mathsf A} t'} 
\bm {\tilde N}(t') dt'.
\ee
Approximating $\bm {\tilde N}(t')$ by the leading terms of the Taylor
expansion
about the lower limit $t$ of the integral in  (\ref{eq:solvV}) followed  
by the variable transformations $ t' \rightarrow \tau' + t$ and  subsequently
 $ \tau' = \tau dt $ one arrives at:
\bee 
\label{eq:solvVn}
\bm V(t+dt) = e^{{\mathsf A} dt} \bm V(t) - dt \int_0^{1} e^{{(1- \tau) \mathsf A}dt } \,\left[\bm{ \tilde N}(t)   + \tau \,  dt\, \frac{\bm{\tilde  N} (t) - \bm{\tilde N}(t-dt)}{dt} \right]d\tau .
\ee

In \cite{koikari2009}, one finds for an arbitrary matrix $\mathsf{M}$, the
following definition of the matrix functions $\phi_k(\mathsf{M})$:
\bee 
\label{eq:phiA}
\phi_0(\mathsf{M}) =  e^{\mathsf{M} }, \quad  \phi_k(\mathsf{M}) =
\frac{1}{(k-1)!} \int_0^1   e^{(1-\tau) \mathsf{M}} \; \tau^{k-1} d \tau; \quad
k =1, 2  \dots.
\ee
Thus, (\ref{eq:solvVn}) can be rewritten as:
\bee 
\label{eq:solvphi1}
\bm V(t+dt) = \phi_0({\mathsf A}\, dt)\,  \bm V(t) -  
\,  dt\, \phi_1({\mathsf A}\, dt) \,\bm{\tilde N}(t)  - dt^2\, \phi_2({\mathsf A}\, dt)\, \left(\frac{ \bm{\tilde N}(t)  - \bm{\tilde N}(t-dt)}{dt}\right) .  
\ee

Note, that both $\bm V$ and the secondary meanflow $\bm U$ (see 
\ref{eq:meanU})) are calculated using the time exponential method.

The time stepping scheme described in (\ref{eq:solvphi1}) incorporates the convergence to stationary solutions $\bm V_s$ of (\ref{eq:gl1n}) which have to fulfill  ${\mathsf A} \bm V_s - \bm{\tilde N}_s = 0 $. This can be proven using the recurrence identities of the matrix  operators  $\phi_k(\mathsf{M})$ as
\bee
\label{eq:phirecur}
\phi_k(\mathsf{M})=\mathsf{M}^{-1}[\phi_{k-1}(\mathsf{M})-I]\,.
\ee

%%%%%%%%%%%%%%%%%%%%%%%%%%%%%%%%%%%%%%%%%%%%%%%%%%%%%%%%%%%%%%%%%%%%%%%%%%%%%%%%%%%%%%%%%%%%
 
It should be remarked that the matrix exponentials could be also treated by
using
a spectral representation of  ${\mathsf A}$   (\ref{eq:gl1n})  in
terms of its direct and
adjoint eigenfunctions.
In this way, one makes immediate contact to the method used in
\cite{Pesch1996}. While this procedure has been successfully applied in a
series of papers on complex patterns in standard RBC \citep{Bodenschatz2000, Egolf2000}, its
application to ILC requires particular care and is less robust, since the
spectral properties of the operator  ${\mathsf A}$  are
complicated  \citep{rudakov1967,Chen1989}.

%%%%%%%%%%%%%%%%%%%%%%%%%%%%%%%%%%%%%%%%%%%%%%%%%%%%%%%%%%%%%%%%%%%%%%%%%%%%%%%%%%%

\section{Additional remarks  on the linear stability calculations}
\label{App:clin}

In \S\ref{sec:rolls}, the linear instability of the basic state, against rolls with wavevector $\bm q = q(\cos \psi , \sin \psi)$ was discussed. This case corresponds to the existence of an eigenvalue $\sigma$
of (\ref{eq:eig1}) with $Re[\sigma] \gtrsim 0]$. In the following, we will exclusively concentrate on the   $\theta, f$ components   obtained from Fourier transformation of (\ref{eq:incfgh})), since $\Re[\sigma] < 0$ for eigenvalues of the separated  $\Phi-$  equation.  Since (\ref{eq:incfgh}) are invariant against the transformations $ (x,z)  \rightarrow -(x,z)$  and separately   $ (y  \rightarrow  -y)$, it is sufficient to restrict $\psi$ to the interval $(0  \le \psi < 90^{\circ})$. As already mentioned in \S\ref{sec:rolls}, and documented in figure \ref{fig:2} for $\Pran = 1.07$, it is sufficient to investigate only the special cases either $\psi = 90^{\circ}$ (longitudinal rolls) or $\psi = 0^{\circ}$
(transverse rolls). A proof  can be found in \cite{Gershuni1969}.  In subsection  \ref{App:cobl}, we will present our own very short version.

\subsection{Linear stability results for small and large $\Pran$}
\label{subsec:appapran}
%As already mentioned in \S\ref{sec:rolls}, a comprehensive overview and discussion of  the linear stability of the basic state for arbitrary $\Pran$ can be found in \cite{Chen1989}. 
We have reproduced some of the earlier results
in the literature as validation of our numerical methods.
In general, the codimension 2  point $\gamma=\gamma_{c2}$, where the
critical Rayleigh numbers $R^l_c(\gamma)$  of the longitudinal rolls and
$R^{t}_c(\gamma)$ of the transverse ones are equal, moves continuously
towards  $\gamma=0$  for decreasing $\Pran$ . Below $\Pran <  0.264$  (more precisely
calculated in \cite{Fujimura1992}), only the primary bifurcation to transverse rolls
prevails even at incremental inclinations from the flat $\gamma=0$ state. 

On the other hand, for $ \Pran>1.07$, the codimension 2
point moves continuously towards $\gamma=90^0$ and eventually the  bifurcation to transverse rolls is  relevant only in the range $90^{\circ}\lesssim\gamma<180^{\circ}$. In addition, the  primary bifurcation to transverse rolls becomes oscillatory at 
large $\Pran$. For the vertical case ($\gamma=90^{\circ}$) we calculate that this happens at
$\Pran\gtrsim12.45$, in agreement with \cite{Fujimura1992}.
%(for details see figure 3 in  \cite{Chen1989}).
The   oscillatory  bifurcation has been  perfectly
reproduced by our own calculations. 
At a slightly larger  $\Pran = 12.7$ again at  $\gamma=90^0$, we find $R_c =88220$, $\omega_c=504.6$ which is in excellent  agreement with
\citet{Bergholz1977}.

\subsection{Bifurcation of oblique rolls}
\label{App:cobl}

It is useful  to exploit a certain form invariance of the linear equations
(\ref{eq:linthzet1n}, \ref{eq:incfn}) in Fourier space using  the
transformation $f \rightarrow f/R$. It turns out that   the Rayleigh
number $R$, the inclination angle $\gamma$  and the oblique roll angle $\psi$ 
appear first in the combination $R \sin \gamma \cos \psi$  due to the
contributions of $\bm U_0$ (see (\ref{eq:basest})) and  furthermore in
(\ref{eq:incfn}) as $R \cos \gamma$.  In addition, these
equations depend only on $|\bm q|$.

Thus, the critical 
Rayleigh number $R^l_c(\gamma)$ of the longitudinal rolls  ($\psi =
90^{\circ}$)  is determined by $R_c^l(\gamma) \cos \gamma =  R_c^l(\gamma =0) =
1707. 824$. At the codimension 2 point $\gamma_{c2}$, we have 
$R_c^l(\gamma_{c2}) =  R_c^t(\gamma_{c2})$, where  $R^t(\gamma)$ denotes the
critical Rayleigh number of the transverse rolls. Thus, $\gamma_{c2}$ is
determined by the 
transcendental  equation $R^{t}_c(\gamma_{c2})\cos (\gamma_{c2}) =
 R_{c0}$ which has always a unique  solution $0 < \gamma_{c2} <  90^{\circ}$ for
$0.246 <\Pran < 12.45$.

The form invariance of the $\theta, f$ equations implies a direct relation
between transverse eigensolutions  ($\psi =0$)  of (\ref{eq:eig1}) for
given values of $\gamma \ne 90^{\circ}, |\bm q|, \sigma, \Pran$  and the oblique
eigensolutions ($\psi \ne 90^{\circ}$) for the same values of  $|\bm q|,
\sigma, \Pran$  but  at  a different inclination angle $\Gamma
\ne \gamma$ and for a Rayleigh number $R =R^{t}(\Gamma) \ne R^{ob}$. 
In detail, we have:
\begin{equation}
\label{eq:traf1}
 R^{ob} \sin\gamma \cos\psi =  R^{t}(\Gamma)  \sin\Gamma, \quad
 R^{ob} \cos\gamma  = R^{t} \cos\Gamma  \Longrightarrow
\tan(\gamma)\cos(\psi) = \tan(\Gamma). 
\end{equation}

In the special case $\Gamma=\gamma=90^{\circ}$ equation (\ref{eq:traf1})
simplifies to $R^{ob}\cos\psi=R^{t}$. In \cite{Gershuni1969} the authors have arrived to analogous relations. As a general consequence  an explicit analysis of linear oblique rolls is not necessary, since they can be determined from the transverse rolls on the basis of  (\ref{eq:traf1}).

Equations (\ref{eq:traf1}) are useful to characterize the stationary bifurcation to oblique rolls 
($\sigma =0$) at medium $\Pran$,  which is characterized  by the curve $R =
R^{ob}_c(\gamma,\psi)$ as function of the inclination angle $\gamma$. This curve
 crosses the  longitudinal threshold curve $R_c^l(\gamma)$ at the codimension 2
point $\gamma^{ob}_{c2}(\psi) < 90^{\circ}$, which  is  thus
 determined by the relation $R^{ob}_c(\gamma^{ob}_{c2},\psi)  =
R_{c0}/\cos(\gamma^{ob}_{c2})$.

Thus, (\ref{eq:traf1}) predicts that there exists a certain  angle $\Gamma'$ which allows for expressing $R^{ob}_c(\gamma^{obl}_{c2},\psi)$ by $R^{t}_c(\Gamma')$ as follows:
\begin{equation}
R^{ob}_c(\gamma^{ob}_{c2},\psi)\cos
(\gamma^{ob}_{c2}) = R^{t}_c(\Gamma') \cos
(\Gamma') = R_{c0};\hspace{0.2cm}
\tan(\gamma^{ob}_{c2})\cos(\psi) = \tan(\Gamma').
\label{eq:gamob}
\end{equation}
From the first equation  in  (\ref{eq:gamob}), we conclude  $\Gamma'  = 
\gamma_{c2}$  and from the second one $\gamma^{ob}_{c2} >
\gamma_{c2}$ for ($0< \psi < 90^{\circ}$), such that for  $\gamma >\gamma_{c2}$
pure transverse rolls prevail at onset.  From  (\ref{eq:gamob}), is also obvious that the dip in the
$q_c(\gamma)-$curve  in the transverse case ($\psi =0)$ is mapped to
corresponding ones
for $\psi \ne 0$ with the same $|\bm q_c|$;  this immediately explains their equal heights in figure \ref{fig:4}.

Though not relevant for the onset of convection, the sudden, strong increase of
$R^t_c(\gamma)$ curve  at $\gamma \approx 26.5^{\circ}$ in figure \ref{fig:2}
is interesting. For $\Pran > 1.75$ even  discontinuous jumps in
$R^t_c$ and $q_c^t$ at $\gamma \approx 30^{\circ}$ develop. These are associated
with the phenomenon of two disconnected neutral curves as seen in figure 6 in
\cite{Chen1989}).

\bibliographystyle{jfm} 

\bibliography{literat}

\begin{thebibliography}{35}
\expandafter\ifx\csname natexlab\endcsname\relax\def\natexlab#1{#1}\fi

\bibitem[Bergholz(1977)]{Bergholz1977}
{\sc Bergholz, R.F.} 1977 Instability of steady natural convection in a
  vertical slot. {\em Journal of Fluid Mechanics\/} {\bf 94}, 743--768.

\bibitem[Birikh {\em et~al.\/}(1972)Birikh, Gershuni, Zhukhovitzkii \&
  Rudakov]{birikh1972}
{\sc Birikh, R.~V., Gershuni, G.Z, Zhukhovitzkii, E.M. \& Rudakov, R.~N.} 1972
  On oscillatory instability of plane parallel convective motion in a vertical
  channel. {\em Prikl. Mat. i Mekh. (PMM), 745\/} {\bf 36}.

\bibitem[Bodenschatz {\em et~al.\/}(2000)Bodenschatz, Pesch \&
  Ahlers]{Bodenschatz2000}
{\sc Bodenschatz, E., Pesch, W. \& Ahlers, G.} 2000 Recent developments in
  {R}ayleigh-{B}{\'e}nard convection. {\em Annual Review of Fluid Mechanics\/}
  {\bf 32}, 709--778.

\bibitem[Boyd(2001)]{Boyd2001}
{\sc Boyd, J.P.} 2001 {\em Chebyshev and Fourier Spectral methods\/}. Dover.

\bibitem[de~Bruyn {\em et~al.\/}(1996)de~Bruyn, Bodenschatz, Morris, Trainoff,
  Hu, Cannell \& Ahlers]{deBruyn1996}
{\sc de~Bruyn, J.~R., Bodenschatz, E., Morris, S.~W., Trainoff, S.~P., Hu, Y.,
  Cannell, D.~S. \& Ahlers, G.} 1996 Apparatus for the study of
  {R}ayleigh-{B}\'enard convection in gases under pressure. {\em Review of
  Scientific Instruments\/} {\bf 67}, 2043--2067.

\bibitem[Busse(1989)]{Busse1989}
{\sc Busse, F.~H.} 1989 Fundamentals of thermal convection. In {\em Mantle
  Convection: Plate Tectonics and Global Dynamics\/} (ed. W.H. Peltier).
  Montreaux: Gordon and Breach.

\bibitem[Busse \& Clever(1979)]{Busse1979}
{\sc Busse, F.~H. \& Clever, R.~M.} 1979 Instabilities of convection rolls in a
  fluid of moderate {P}randtl number. {\em Journal of Fluid Mechanics\/} {\bf
  91}, 319--335.

\bibitem[Busse \& Clever(1992)]{Busse1992}
{\sc Busse, F.~H. \& Clever, R.~M.} 1992 Three-dimensional convection in an
  inclined layer heated from below. {\em Journal of Engineering Mathematics\/}
  {\bf 26}, 1--49.

\bibitem[Busse \& Clever(1996)]{Busball1996}
{\sc Busse, F.~H. \& Clever, R.~M.} 1996 The sequence-of-bifurcations approach
  towards an understanding of complex flows. In {\em Mathematical Modelling and
  Simulation in Hydrodynamic Stability\/} (ed. D.~N. Riahi). World Scientific,
  Singapore.

\bibitem[Busse \& Clever(2000)]{Busse2000}
{\sc Busse, F.~H. \& Clever, R.~M.} 2000 Bursts in inclined layer convection.
  {\em Physics of Fluids\/} {\bf 12}, 2137--2140.

\bibitem[Chandrasekhar(1961)]{chandra1961}
{\sc Chandrasekhar, S.} 1961 {\em Hydrodynamic and Hydromagnetic Stability\/}.
  Clarendon Press, Oxford.

\bibitem[Chen \& Pearlstein(1989)]{Chen1989}
{\sc Chen, Y.~M. \& Pearlstein, A.~J.} 1989 Stability of free-convection flows
  of variable-viscosity fluids in vertical and inclined slots. {\em Journal of
  Fluid Mechanics\/} {\bf 198}, 513--541, note that the inclination angle
  ($\delta$ in this work) is measured with respect to the vertical direction.

\bibitem[Clever \& Busse(1977)]{Clever1977}
{\sc Clever, R.~M. \& Busse, F.~H.} 1977 Instabilities of longitudinal
  convection rolls in an inclined layer. {\em Journal of Fluid Mechanics\/}
  {\bf 81}, 107--127.

\bibitem[Clever \& Busse(1995)]{Busse1995}
{\sc Clever, R.~M. \& Busse, F.~H.} 1995 Tertiary and quarternary solutions for
  convection in a vertical fluid layer heated from the side. {\em Chaos,
  Solitons \& Fractals\/} {\bf 5}, 1795--1803.

\bibitem[Cross \& Hohenberg(1993)]{CroHo1993}
{\sc Cross, M.C. \& Hohenberg, P.C.} 1993 Pattern formation outside of
  equilibrium. {\em Rev. Mod. Phys.\/} {\bf 65}, 852--1111.

\bibitem[Daniels(2002)]{karenthesis2002}
{\sc Daniels, K.} 2002 Pattern formation and dynamics in inclined layer
  convection. PhD thesis, Cornell University, USA.

\bibitem[Daniels \& Bodenschatz(2002)]{Karen2002}
{\sc Daniels, K.~E. \& Bodenschatz, E.} 2002 Defect turbulence in inclined
  layer convection. {\em Phys. Rev. Lett.\/} {\bf 88}, 034501.

\bibitem[Daniels {\em et~al.\/}(2008)Daniels, Brausch, Pesch \&
  Bodenschatz]{Karen2008}
{\sc Daniels, K.~E., Brausch, O., Pesch, W. \& Bodenschatz, E.} 2008
  Competition and bistability of ordered undulations and undulation chaos in
  inclined layer convection. {\em Journal of Fluid Mechanics\/} {\bf 597},
  261--282.

\bibitem[Daniels {\em et~al.\/}(2000)Daniels, Plapp \& Bodenschatz]{Karen2000}
{\sc Daniels, K.~E., Plapp, B.~B. \& Bodenschatz, E.} 2000 Pattern formation in
  inclined layer convection. {\em Phys. Rev. Lett.\/} {\bf 84}, 5320--5323.

\bibitem[Daniels {\em et~al.\/}(2003)Daniels, Wiener \& Bodenschatz]{Karen2003}
{\sc Daniels, K.~E., Wiener, R.~J. \& Bodenschatz, E.} 2003 Localized
  transverse bursts in inclined layer convection. {\em Phys. Rev. Lett.\/} {\bf
  91}, 114501.

\bibitem[Dominguez-Lerma {\em et~al.\/}(1984)Dominguez-Lerma, Ahlers \&
  Cannell]{Ahl1984}
{\sc Dominguez-Lerma, M.~A., Ahlers, G. \& Cannell, D.S.} 1984 Marginal
  stability curve and linear growth rate for rotating couette-taylor flow and
  rayleigh.b´enard convection. {\em Phys. Fluids\/} {\bf 27}, 856--860.

\bibitem[Egolf {\em et~al.\/}(2000)Egolf, Melnikov, Pesch \& Ecke]{Egolf2000}
{\sc Egolf, D., Melnikov, I.V., Pesch, W. \& Ecke, R.} 2000 Extensive
  spatiotemporal chaos in {R}ayleigh-{B}{\'e}nard convection. {\em Nature\/}
  {\bf 404}, 733.

\bibitem[Fujimura \& Kelly(1992)]{Fujimura1992}
{\sc Fujimura, K. \& Kelly, R.E} 1992 Mixed mode convection in an inclined
  slot. {\em Journal of Fluid Mechanics\/} {\bf 246}, 545--568.

\bibitem[Gershuni \& Zhukhovitzkii(1969)]{Gershuni1969}
{\sc Gershuni, G.Z \& Zhukhovitzkii, E.M.} 1969 Stability of plane-parallel
  convective motion with respect to spatial perturbations. {\em Prikl. Mat. i
  Mekh. (PMM), 855\/} {\bf 33}.

\bibitem[Hart(1971)]{Hart1971}
{\sc Hart, J.~E.} 1971 Stability of flow in a differentially heated inclined
  box. {\em Journal of Fluid Mechanics\/} {\bf 91}, 319--335.

\bibitem[Koikari(2009)]{koikari2009}
{\sc Koikari, S.} 2009 Planar measurements of differential diffusion in
  turbulent jets. {\em ACM Transactions on Mathematical Software\/} {\bf 36},
  12.

\bibitem[Lappa(2009)]{Lappa2009}
{\sc Lappa, M.} 2009 {\em Thermal Convection, Patterns, Evolution and
  Stability\/}. Wiley.

\bibitem[Lemoult {\em et~al.\/}(2014)Lemoult, Gumowski, Aider \&
  Wesfreid]{Lemoult2014}
{\sc Lemoult, G., Gumowski, K., Aider, J-L. \& Wesfreid, J.E.} 2014 Turbulent
  spots in channel flow: An experimental study. {\em The European Physical
  Journal E\/} {\bf 37}~(4), 25.

\bibitem[Pesch(1996)]{Pesch1996}
{\sc Pesch, W.} 1996 Complex spatiotemporal convection patterns. {\em Chaos: An
  Interdisciplinary Journal of Nonlinear Science\/} {\bf 6}, 348--357.

\bibitem[Rudakov(1967)]{rudakov1967}
{\sc Rudakov, R.~N.} 1967 Spectrum of perturbations and stability of convective
  motion between vertical planes. {\em Prikl. Mat. i Mekh. (PMM), 349\/} {\bf
  31}.

\bibitem[Ruth {\em et~al.\/}(1980)Ruth, Hollands \& Raithby]{Ruth1980}
{\sc Ruth, D.W., Hollands, K. G.~T. \& Raithby, G.~D.} 1980 On free convection
  experiments in inclined air layers heated from below. {\em Journal of Fluid
  Mechanics\/} {\bf 96}, 461--479.

\bibitem[Swinney \& Gollub(1985)]{swin85}
{\sc Swinney, H.L. \& Gollub, J.P.} 1985 {\em Hydrodynamic Instabilities and
  the Transition to Turbulence, Second Edition\/}. Springer-Verlag Berlin.

\bibitem[Trainoff \& Canell(2002)]{Trainoff2002}
{\sc Trainoff, S.~P. \& Canell, D.~S.} 2002 Physical optics treatment of the
  shadowgraph. {\em Phys. Fluids\/} {\bf 14}, 1340--1363.

\bibitem[Tuckerman {\em et~al.\/}(2014)Tuckerman, Kreilos, Schrobsdorff,
  Schneider \& Gibson]{Tuckerman2014}
{\sc Tuckerman, L.~S., Kreilos, T., Schrobsdorff, H., Schneider, T.~M. \&
  Gibson, J.~F.} 2014 Turbulent-laminar patterns in plane poiseuille flow. {\em
  Physics of Fluids\/} {\bf 26}~(11), 114103.

\bibitem[Vest \& Arpaci(1969)]{Vest1969}
{\sc Vest, C.~M \& Arpaci, V.S.} 1969 Stability of natural convection in a
  vertical slot. {\em Journal of Fluid Mechanics\/} {\bf 36}, 1--15.

\end{thebibliography}

\end{document}